\documentclass[smallextended]{svjour3}       
\smartqed  
\usepackage{csl}
\usepackage{graphicx}
\usepackage{bm}
\usepackage{amsmath,amsfonts,amssymb,amscd,amstext}
\usepackage{xcolor}
\usepackage{svg}
\usepackage[plain]{algorithm2e}
\usepackage{subcaption}
\usepackage{lipsum}
\usepackage{caption}
\usepackage{booktabs}
\usepackage[utf8]{inputenc} 
\usepackage[T1]{fontenc}
\usepackage[sort]{cite}
\usepackage[colorinlistoftodos]{todonotes}
\usepackage{nomencl}
\makenomenclature

\newlength{\imagewidth}

\newcommand{\subgraphics}[2]{
\settowidth{\imagewidth}{\includegraphics[width=3.3cm]{#1}}%
\begin{subfigure}{\imagewidth}%
    \includegraphics[width=3.3cm, clip]{#1}%
    \caption{#2}%
\end{subfigure}%
}

\begin{document}

\csltitle{Simultaneous Optimal System and Controller Design for Multibody Systems with Joint Friction using Direct Sensitivities}
\cslauthor{Adwait Verulkar,
        Corina Sandu,
        Adrian Sandu, and
        Daniel Dopico}
\cslyear{23}
\cslreportnumber{7}
\csltitlepage

\title{Simultaneous Optimal System and Controller Design for Multibody Systems with Joint Friction using Direct Sensitivities}

\titlerunning{Optimization of Multibody Systems with Friction}        

\author{Adwait Verulkar\textsuperscript{1}\and
        Corina Sandu\textsuperscript{1} \and
        Adrian Sandu\textsuperscript{2} \and
        Daniel Dopico\textsuperscript{3}}


\institute{
\textsuperscript{1}Terramechanics, Multibody and Vehicle Systems Laboratory,\\
Department of Mechanical Engineering,\\
Virginia Tech, Blacksburg, United States. \email{[adwaitverulkar, csandu]@vt.edu}\\  
\textsuperscript{2}Computational Science Laboratory, Department of Computer Science,\\
Virginia Tech, Blacksburg, United States. \email{asandu7@vt.edu}\\ 
\textsuperscript{3}Laboratorio de Ingeniería Mecánica, Department of Naval and Industrial Engineering, \\
University of A Coruña, A Coruña, Spain. \email{ddopico@udc.es}
 }

\date{Received: / Accepted: date}

\maketitle
\begin{abstract}
Real-world multibody systems are often subject to phenomena like friction, joint clearances, and external events. These phenomena can significantly impact the optimal design of the system and its controller. This work addresses the gradient-based optimization methodology for multibody dynamic systems with joint friction using a direct sensitivity approach for gradient computation. After a thorough review of various friction models developed over the years, the Brown McPhee model has been found to be the most suitable for the study due to its accuracy for dynamic simulation and its compatibility with sensitivity analysis. The methodology supports co-design of the system and its controller, which is especially relevant for applications like robotics and servo-mechanical systems where the actuation and the design are highly dependent on each other. Numerical results are obtained using a new implementation of the MBSVT (Multi-Body Systems at Virginia Tech) software package; MBSVT 2.0 is reprogrammed in Julia for ease of implementation while maintaining high computational efficiency. Three case studies are provided to demonstrate the attractive properties of simultaneous optimal design and control approach for certain applications.

\end{abstract}
\newpage
\section{Introduction}
\label{intro}
Performance analysis of mechanical systems is usually conducted by studying the dynamics using a simulation software. Due to advances in differential equation solvers and optimization techniques, it has become increasingly convenient to perform sensitivity analysis and gradient-based dynamic optimization on large scale dynamical systems involving multiple design/control parameters. Multibody systems are a special type of dynamic systems that consist of multiple links connected by joints. This gives rise to relatively complex systems of equations that require special solvers. It is also possible to simulate realistic behaviors in these systems due to recent research on multibody systems with friction \cite{Marques2019,Fraczek2011,Pennestri2007,Haug2018b}, clearances \cite{Tian2018,Flores2004,Flores2006,Orden2005}, events \cite{Corner2019,Corner2020}, and even a combination of these \cite{FLORES2023105305}. Such behaviors further add to the complexity of the resulting system of equations. 

When addressing optimal design and control in dynamical systems, sensitivity analysis and, consequently, gradient-based optimization emerge as leading methodologies due to their notable efficiency and speed, setting them apart from gradient-free search methods like Nelder-Mead \cite{Nelder1965}, as well as evolutionary optimization techniques such as the genetic algorithm \cite{Holland1992} and particle swarm optimization \cite{Kennedy1995}. Direct sensitivity methods are applied when the number of free variables is relatively small in comparison with the objective functions, while adjoint sensitivity methods tend to be more efficient when the number of free variables is large \cite{Dopico2015,Alvaro_ALI3P_direct_sensitivity,Maciag2020_hamiltonian_sensitivity}. However, when these techniques are applied to multibody systems which attempt simulation of realistic behavior, like friction and events, several considerations need to be made in terms of solution techniques, accuracy and stability of the sensitivities, and computational efficiency. Optimization is inherently iterative in nature, requiring the dynamics and sensitivity equations to be repeatedly solved. What makes this process especially challenging when it comes to multibody systems with friction, is the implicit and stiff nature of the dynamic and sensitivity equations involved. Hence it becomes imperative to develop a generic and computationally efficient optimization methodology to skillfully tackle these complexities. 

Many dynamic systems rely on active control to achieve specific trajectory goals. In the realm of non-linear optimal control, it's common practice to parameterize the control function. One popular approach involves using basis functions along with their corresponding coefficients as control parameters \cite{PIKULINSKI2021104473,verulkaroptimal}. In contrast, state/output feedback control typically involves a fixed number of control parameters. Frequently, systems must be customized for particular applications, necessitating optimization in both design and control aspects. Recent research has underscored the advantages of simultaneous optimization over separate design and control optimization. This approach, also known as co-design, has the potential to yield more efficient systems by exploiting the flexibility in design to find efficient control solutions compared to fixed design approaches. When the optimization methodology treats design and control parameters on equal footing, it can harness this potential afforded by co-design. The following references highlight the significance of simultaneous design and control in specific applications.

\begin{itemize}
    \item {In certain aerospace applications where there are extreme power to weight concerns it might be beneficial to do simultaneous optimization of control and design \cite{simultaneousDC1}. In this work, co-design is shown to be helpful for the integration of energy management optimization along with optimal vehicle sizing for a hybrid electric propulsion aircraft.}
    \item {In civil engineering applications, building structures for earthquake resistance can be done through simultaneous design and control \cite{Alavi2021}. This type of structure uses active controls to prevent structural damage due to seismic activity.}
    \item {Specifically talking about multibody systems with friction, co-design optimization strategies are also used for design of active suspension systems \cite{Allison2014CodesignOA}. This method is exactly the same as using basis functions to convert a continuous control into discrete parameterizations, referred to as direct transcription.}
\end{itemize}

Simultaneous design and control is especially useful in legged robotic systems. Earlier robots employed high-gain feedback and therefore used considerable joint torque to cancel out the natural dynamics of the machine to follow a desired trajectory. Optimizing the control without making any design considerations may lead to sub-optimal design choices and incurring a very stiff penalty. An example of such a system is ASIMO which is a legged-robot that uses roughly 20 times the energy (scaled) that a human uses to walk on a flat surface, as measured by the cost of transport \cite{Collins2005}. Leveraging the design and the dynamics to suit application's control needs is crucial for such systems \cite{7041347}.

Recent advances in Automatic Differentiation (AD) have started a discussion on discrete versus continuous sensitivity analysis methods \cite{RevelsLubinPapamarkou2016_ForwardADJulia,Alfonso2013_dynamic_response_optimization_using_AD,Ma2021_comparison_of_AD_in_DE}. Since the 1990s, Automatic Differentiation (AD) has been a popular tool for sensitivity computation in ODEs/DAEs, and, by extension, for multibody systems as well \cite{Petzold2006,FEEHERY199741,LI2000131,ADIFOR}. Conventionally, it has been a practice to differentiate the dynamic equations of motion to obtain the sensitivity equations. The sensitivity equations obtained through direct differentiation of system dynamics are also known as Tangent Linear Models (TLMs). These sensitivity equations also turn out to be differential equations which can be integrated to obtain the model sensitivities with respect to the design or control parameters. However, in recent times, another course of action is to simply perform AD of the differential equation solver step itself \cite{ADofDEsolvers}. The discrete sensitivity approach is mathematically equivalent to the previous continuous approach. However, in certain cases, it may offer improved ease of programming without sacrificing computational efficiency. The main reason being the ability of AD to perform compiler optimizations by making use of structure between the primal and derivative constructions. This allows the AD code to perform Single Instruction on Multiple Data (SIMD), constant folding and Common Subexpression Elimination (CSE) \cite{Muchnick1997}. An experienced user can program these optimizations manually. However, there lies a trade-off between the time and effort required to obtain an efficient code versus the computational cost saved. Additionally, there are differences in stability of continuous versus discrete approaches for reverse-mode AD, as denoted in \cite{rackauckas2021universal,Kim_2021}. It is important to note, however, that AD might not be a viable option in cases where the dynamic equations have to be implicitly solved. The iterative convergence required in implicit differential equation solvers creates unnecessary computations that can be avoided in continuous approaches by leveraging the implicit function theorem. Also, AD of implicit solvers can only yield approximate derivative since the convergence is solved to a certain tolerance. A good application of AD is in building the derivative components of continuous sensitivity methods. These components can be obtained without computing a full Jacobian through efficient Jacobian-vector products. In this work, careful considerations have been taken to ensure efficiency by using a hybrid differentiation approach. For multibody systems, most Jacobians and Jacobian-vector products required in dynamics and sensitivity computations are available as a closed-form expressions that are simplified to a large extent using domain-specific mathematical identities. This makes the derivative computation efficient to the point that it can be represented by a single non-allocating function call. Expressions involving Jacobians and Jacobian-vector products with respect to states can therefore be more efficiently computed through Manual Differentiation (MD), whereas those involving design or control parameters require AD. This not only alleviates the user's responsibility for providing Jacobian or Jacobian-vector products for specific multibody systems, but also allows improved space complexity and memory allocations. Symbolic Differentiation (SD) is another popular approach, which often exhibits similar time and space complexity to that of MD. However, most computer algebra systems do not have the capability to carry out expression simplifications in a sophisticated way that will take advantage of domain specific mathematical identities. This frequently results in complicated expressions, often referred to as expression swell, for higher order derivatives and matrix calculus. Symbolic code frequently contains several auxiliary coefficients which add to the computational cost. Also, some expressions and their derivatives cannot be generated through symbolic computations like those involving large matrix inversions due to the prohibitively slow and memory intensive nature of symbolic computations. Symbolic differentiation also fails in cases where a function may not be represented by a mathematical formula and therefore cannot handle complex control flow further limiting its expressivity \cite{Griewank2008,baydin2018automatic}. Finally, there is the question of automated translation of a symbolic expression to an efficient numerical function. The user seldom has complete control over this process and may lead to suboptimal code. AD can handle all of these edge cases while the user being in greater control of the code. 

Table \ref{tab:comparison} highlights the differences in computational cost using various differentiation approaches for some common terms required in multibody formulations such as the constraint Jacobian $\bm\Phi_\mathbf{q}$, and various Jacobian-vector products like the total time derivative of the constraint vector $\bm\Phi_{\mathbf{q}}\dot{\mathbf{q}}$, acceleration term $(\bm\Phi_\mathbf{q}\dot{\mathbf{q}})_{\mathbf{q}}\dot{\mathbf{q}}$ and the generalized reaction term $\bm\Phi_{\mathbf{q}}^\text{T}\bm\lambda$. The matrices are computed for the slider-crank case study discussed in Section \ref{sec:test-slider-crank}, which has a total of 20 constraints and 21 generalized coordinates. As it can be observed, MD performs best in terms of both space and time complexity when compared to AD and SD approaches. The speed comes from the fact that analytical Jacobian is computed by effectively a single function call (for sub-Jacobian computation) as opposed to the recursive function calls in the case of AD. However, AD can be made substantially more efficient by employing sparsity detection and matrix coloring techniques \cite{McCourt2015,Alappat2020,gowda2019sparsity} not employed in this analysis.

\begin{table}[ht]
  \centering
  \caption{Cost of different differentiation approaches for computing Jacobians of the slider-crank system discussed in Section \ref{sec:test-slider-crank}}
  \label{tab:comparison}
  \begin{tabular}{lcccc}
    \toprule
    Operation & Manual & Automatic & Symbolic \\
    \midrule
    $\bm\Phi_\mathbf{q}$ & 21.992 $\mu$s / 23.59 kB & 177.768 $\mu$s / 829.34 kB & 1.042 ms / 101.12 kB \\
    $\bm\Phi_\mathbf{q}\dot{\mathbf{q}}$ & 16.560 $\mu$s / 6.15 kB & 152.169 $\mu$s / 224.50 kB & 858.22 $\mu$s / 32.26 kB \\
    $\bm\Phi_{\mathbf{q}}^\text{T}\bm\lambda$ & 30.254 $\mu$s / 15.22 kB & 178.494 $\mu$s / 122.28 kB & 995.59 $\mu$s / 38.26 kB \\
    $(\bm\Phi_\mathbf{q}\dot{\mathbf{q}})_{\mathbf{q}}\dot{\mathbf{q}}$ & 18.223 $\mu$s / 20.79 kB & 159.108 $\mu$s / 270.58 kB & 938.31 $\mu$s / 40.66 kB \\
    \bottomrule
  \end{tabular}
\end{table}

This paper presents the methodology for direct sensitivity analysis and dynamic optimization of multibody systems with friction. By converting the continuous control signal into a parameterized form, the methodology can be applied to co-design case studies as well. Direct sensitivity approach for gradient computation is efficient and outperforms reverse-mode (adjoint) approach for optimization problems involving a \textit{relatively small} number of design/control parameters in comparison to the number of state variables. The formulation uses centroidal body-fixed reference frames with the orientation of the bodies defined using Euler parameters. The resulting matrices involved in the equations of motion are sparse, thereby allowing for efficient solution techniques such as Newton-Krylov \cite{NewtonKrylov},  Generalized Minimal RESidual Method (GMRES) \cite{GMRES2005}, and Jacobian-free Newton Krylov \cite{KNOLL2004357}. Three case studies are provided that apply the methodology for a pure control optimization example of inverted pendulum, a pure design optimization example of a governor mechanism, and a co-design example of a spatial slider-crank mechanism. The results are obtained using a revised version of MBSVT \cite{MBSVT_IDETC2014} reprogrammed in Julia. The original software was written in Fortran 2003 with capabilities for kinematic and dynamic simulation of multibody systems, direct and adjoint sensitivity analysis, and gradient-based optimization. A key drawback of MBSVT was the use of full algebra formulations which are inefficient for reference point coordinates with Euler parameters. MBSVT 2.0 takes advantage of recent advances in open-source Julia software such as differential equations solvers (\texttt{DifferentialEquations.jl}), optimization packages (\texttt{Optim.jl}), and support for native forward and reverse mode AD (\texttt{ForwardDiff.jl}, \texttt{Zygote.jl}). Important functions for sensitivity analysis such as Jacobian matrices with respect to design variables, Jacobian-vector products, and gradients of objective functions are computed automatically and need not be user-provided. All derivative terms and sensitivity computations have been thoroughly validated using complex finite differences \cite{Verulkar2022}. Moreover, due to native AD capability, the derivatives with respect to design parameters can be computed by the software (were initially required to be user-provided). Despite being a relatively new programming language, Julia is rapidly gaining popularity within the scientific community for several compelling reasons. Its syntactical convenience, robust computational speed, and the myriad of packages dedicated to scientific computing and machine learning make it an attractive choice for researchers and practitioners alike. MBSVT offers greater flexibility to its users since multibody packages such as MSC Adams, SIMPACK, LMS VirtualLab Motion, RecurDyn, and Simscape are not open-source. Moreover, most of these packages focus on kinematics and dynamics capabilities, with sensitivity analysis and optimization taking the backstage. JModelica/Optimica (now known as Modelon OCT) \cite{AKESSON20101737}, JuliaSim \cite{rackauckas2021composing}, CasADi/IPOPT \cite{Andersson2019,IPOPT2006} are among the few tools that make substantial strides in this regard. These packages employ computational graph based AD to extract gradient information for sensitivity analysis and optimization of dynamic systems. They are excellent packages capable of optimizing multibody systems but require some development effort on the user's end. The revised MBSVT has been planned as an open-source package, thereby making it more accessible in comparison to the previously discussed proprietary alternatives. This release will not only provide access to both, classical and newer multibody formulations, but it will also empower its users with the capability to simulate complex phenomena such as joint friction, events, and clearances. Moreover, Julia's ease-of-development features, such as a built-in package manager, benchmarking tools, well-documented libraries, active user forums, cross-platform compatibility and compilation support, significantly facilitate contributions from new users.

Other Julia libraries that can be employed for multibody dynamics and optimization are the acausal modeling package \texttt{Modia.jl} \cite{dlr144872} and the package \texttt{RigidBodyDynamics.jl} \cite{rigidbodydynamicsjl}. These tools were not explored by the authors in this work, but have been provided as alternatives that could potentially be used for optimization of multibody systems. \texttt{ModelingToolkit.jl}\cite{ma2021modelingtoolkit} is another acausal system model package that can be considered for this study. However, it currently lacks support for the type of differential-algebraic formulations used in this work. It also relies on \texttt{Symbolics.jl} \cite{gowda2021high} for differentiation and sparsity detection, necessitating symbolic traversability of all functions used to model the system.

The novel contributions of this paper are summarized below.
\begin{enumerate}
  \item The work develops a direct-sensitivity based optimization approach for multibody dynamic systems with Brown-McPhee joint friction.
  \item A co-design methodology for simultaneous optimal design of system and controller parameters is introduced. The optimal design problem is constrained by the system dynamics with joint friction.
  \item The methodology's effectiveness is demonstrated through real-world case studies and validated the approach through numerical results.
  \item The MBSVT 2.0 software package has been developed in Julia for sensitivity analysis and optimization of multibody systems. The implementation leverages recent advances in differentiable programming and modern interfaces to differential equation solvers.
\end{enumerate}

The remainder of the paper is organized as follows.
Section \ref{sec:dynamics} reviews the development of equations of motion for multibody systems with friction.
Section \ref{sec:optimization} reviews bound-constrained optimization using direct sensitivity analysis
Section \ref{sec:case_studies} provides the numerical validation of the methodology using various case studies.  
Finally, Section \ref{sec:conclusions} draws conclusions and points to future work.

\section{Dynamics of multibody systems with friction}
\label{sec:dynamics}
This section briefly covers the derivation of the equations of motion for multibody systems with friction using the index-1 DAE formulation. As it will become apparent by the end of this section, the friction forces ultimately depend on the state variables of the equation of motion due to its dependency on the normal force in the joint. This makes the dynamic equations of motion implicit and will require special integration schemes to solve.
\subsection{Joint friction forces}
\label{sec:fric_force}
Computation of generalized friction force for multibody systems is a two step process viz. computation of the magnitude of frictional force and torque at the joint and assembly of the generalized friction force vector. Friction can be modeled using several approaches. The approaches can be segregated based on whether they use a static or a dynamic model. This study uses the Brown and McPhee friction model \cite{Brown2016} to describe joint friction. It is a quasi-static model governed by a single equation. Assuming that most mechanical systems have some lubrication, the friction between surfaces deviates from the dry Coulomb friction model. The mathematical representation of this friction model (excluding viscous friction) is as follows:
\begin{align}
    F_f(v, \bm{\mu}) &= F_n\left[\mu_d\tanh \left(\frac{4v}{v_t}\right)+\frac{\left(\mu_s-\mu_d\right)\left(\frac{v}{v_t}\right)}{\left[\left(\frac{v}{2 v_t}\right)^2+\frac{3}{4}\right]^2}\right].
    \label{eq_S}
\end{align}
The main advantage of this model for sensitivity analysis of multibody systems is its $C^1$ continuity and differentiability, and ability to simulate stiction by allowing relative motion at speeds lower than some user-defined threshold $v_t$. It is important to note that differentiating a friction model may not necessarily yield the same sensitivities as those obtained through piecewise continuous friction models like Coulomb. Haug \cite{Haug2018b} used the Brown and McPhee model for describing the joint friction between two bodies. Determining the transition velocity $v_t$ poses a challenge, and according to Haug \cite{Haug2018b}, it is advisable to set $v_t$ to approximately ten times the average integration time step employed by the solver. A value of $10^{-2}$ to $10^{-3}$ was used in this work.

Another approach in modeling friction is using dynamic friction models. The Gonthier et al. friction model \cite{Gonthier2004} is a more sophisticated model based on the LuGre \cite{LuGre} friction model. Unlike the Brown McPhee model, the Gonthier model incorporates dynamic states that require integration to compute the current friction. A detailed discussion of other friction models that are applicable for sensitivity analysis and optimization has been provided in \cite{Verulkar2022}. 
\subsection{Normal contact forces in joints}
\label{sec_normal_force}
\begin{figure}[htbp]
     \centering
     \begin{subfigure}[b]{0.49\textwidth}
         \centering
         \includegraphics[width=\textwidth]{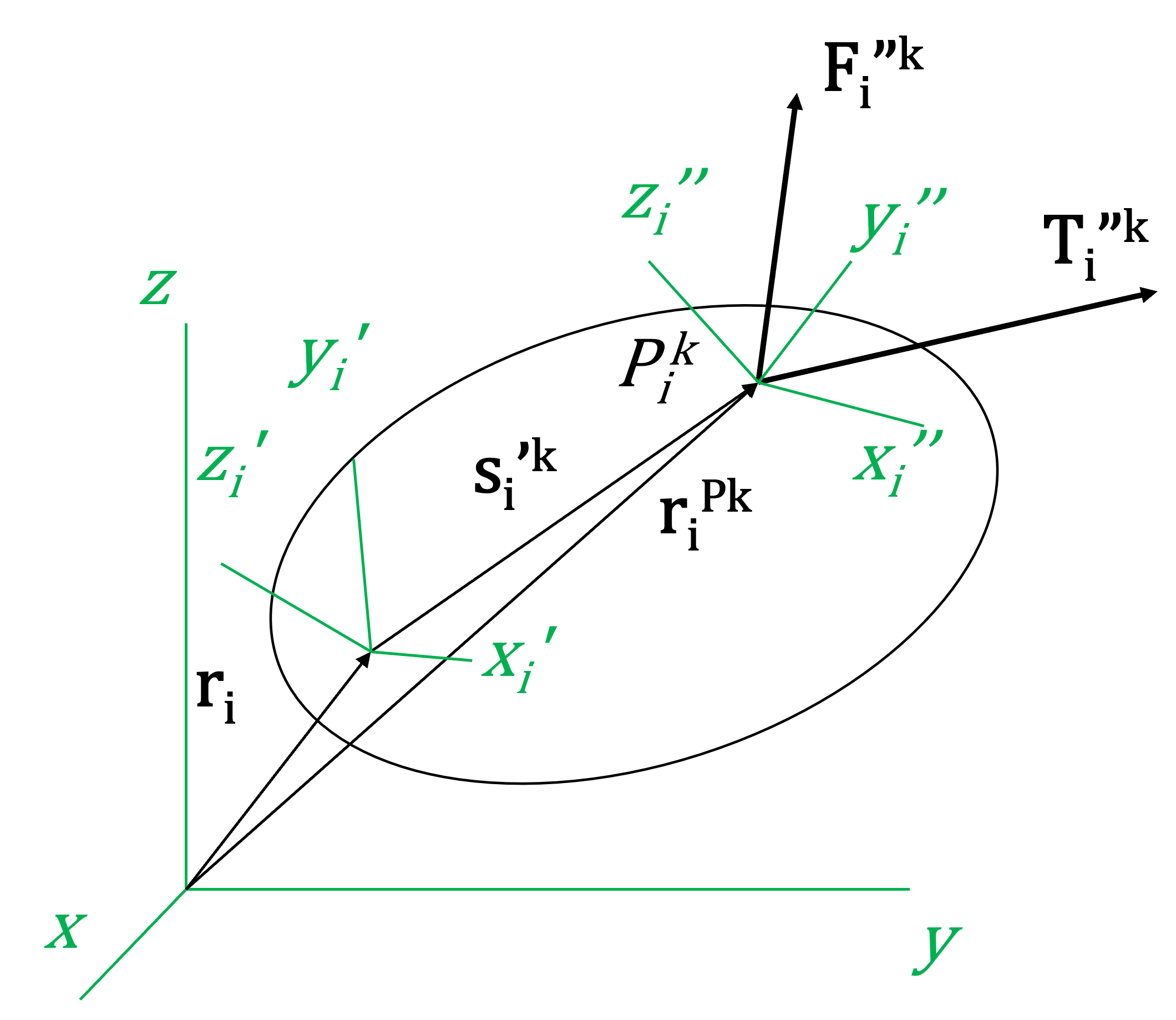}
         \caption{Internal joint forces}
         \label{fig:figure_int_forces}
     \end{subfigure}
     \hfill
     \begin{subfigure}[b]{0.49\textwidth}
         \centering
         \includegraphics[width=\textwidth]{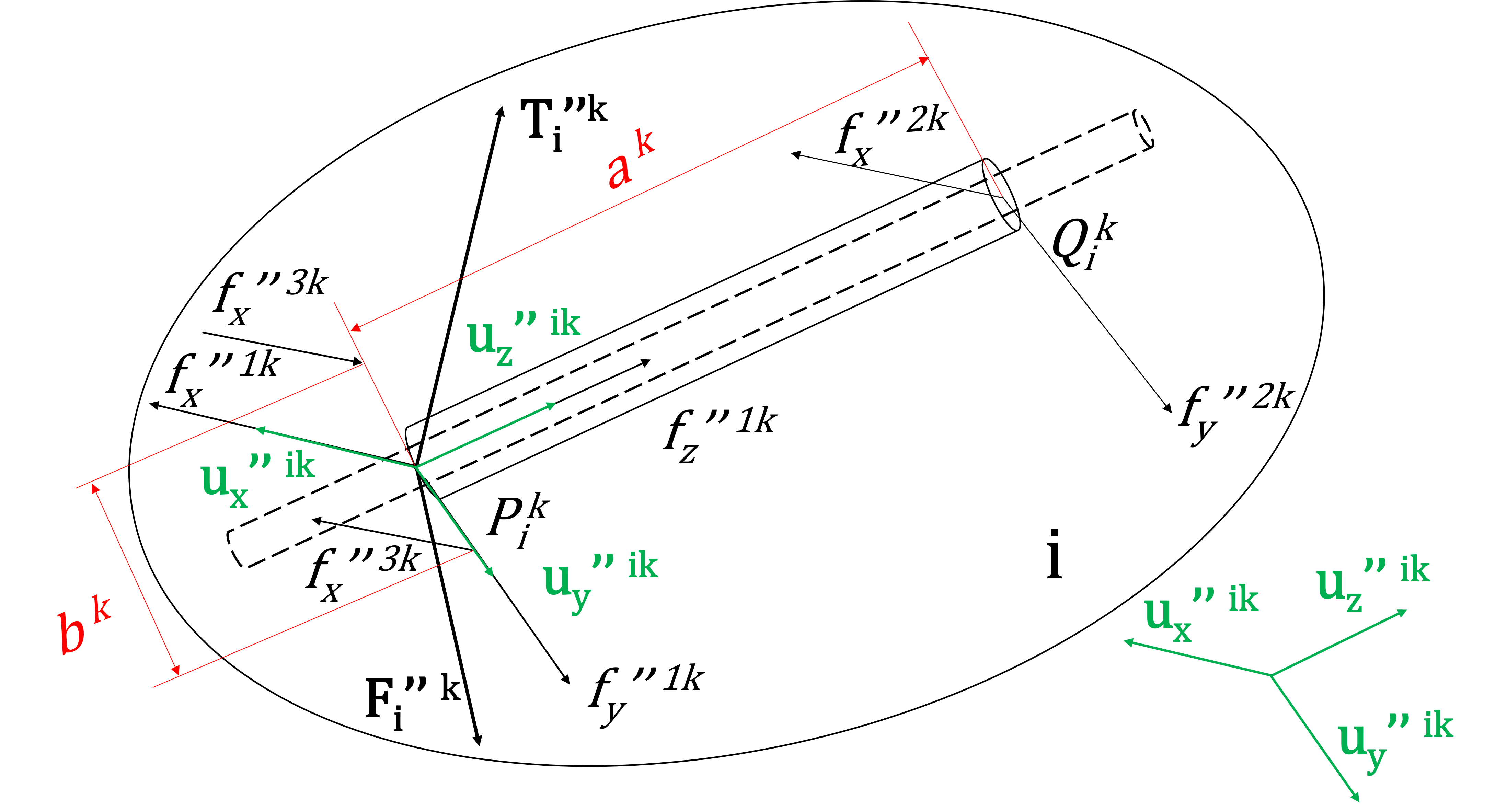}
         \caption{Forces in a cylindrical joint}
         \label{figure_cyl_joint_forces}
     \end{subfigure}
     \hfill
    \caption{Resolution of internal forces in joint reference frame}
    \label{figure_intforce_figures}
\end{figure}

For a multibody system, the constraints are maintained during the motion of the system by internal reaction forces and torques as seen in Figure \ref{fig:figure_int_forces}. The following equation can be used to calculate these physical quantities in the joint reference frame \cite{Haug2018b}:
\begin{align}
    \left\{\begin{array}{c}
    \mathbf{F}_i''^k\\
    \mathbf{T}_i''^k
    \end{array}\right\} =
    \left\{\begin{array}{c}\mathbf{C}_i^{k\text{T}}\,\mathbf{A}_i^\text{T}\,\bm\Phi_{\mathbf{r}_i}^{k\text{T}}\bm\lambda^k\\
\mathbf{C}_i^{k\text{T}}\,\left(\frac{1}{2}\,\mathbf{G}(\mathbf{p}_i)\,\bm\Phi_{\mathbf{p}_i}^{k\text{T}} -\Tilde{\mathbf{s}'}_i^k\,\mathbf{A}_i^\text{T}\bm\Phi_{r_i}^{k\text{T}}\right)\bm\lambda^k
    \end{array}\right\}.
    \label{eq_int_forces} 
\end{align}
In Equation (\ref{eq_int_forces}), for a given body $i$, a holonomic joint $k$ can be defined with the constraints $\bm\Phi^{\text{k}} = \mathbf{0}$. $\mathbf{F}_i''^k$ and $\mathbf{T}_i''^k$ are the reaction force and respectively moments in joint coordinate frame, $\mathbf{C}^k_i$ and $\mathbf{A}_i$ are the joint-to-body and body-to-ground coordinate transformation matrices respectively, $\bm\Phi^{\text{k}}_{\mathbf{r}_i}$ and $\bm\Phi^{\text{k}}_{\mathbf{p}_i}$ are the constraint Jacobians with respect to Cartesian coordinates and Euler parameters of the $i$\textsuperscript{th} body respectively, $\bm\lambda^{\text{k}}$ are the Lagrange multipliers associated with the constraints $\bm\Phi^{\text{k}}$, and $\mathbf{s}'^k_i$ is the position vector in the $i$\textsuperscript{th} body-fixed reference frame for the joint location. To calculate the magnitude of the effective joint normal force, denoted as $F_n$, it is crucial to decompose the forces from the joint reference axes into their components. The joint torque generates a couple within the joint geometry, restricting rotational degrees of freedom and thereby influencing the effective joint normal force.

Haug \cite{Haug2018b} has presented the decomposition of forces and torques in cylindrical, translational, and revolute joints, as depicted in Figure \ref{figure_cyl_joint_forces}. The clearances in the joint required for smooth motion are assumed to be negligible relative to the magnitude of link displacements. The force components have been shown in Figure \ref{figure_intforce_figures}b and can be expressed mathematically using the following equations:
\begin{subequations}
    \begin{align}
        f_x^{\prime\prime 1k} &= -\mathbf{u}_x^{\prime\prime ikT}\mathbf{F}_i''^k+\left(\frac{1}{a^k}\right)\mathbf{u}_y^{\prime\prime ikT}\mathbf{T}_i''^k\text{,} \\
        f_y''^{1k} &= -\mathbf{u}_y''^{ikT}\mathbf{F}_i''^k-\left(\frac{1}{a^k}\right)\mathbf{u}_x''^{ikT}\mathbf{T}_i''^k\text{,}\\
        f_x''^{2k} &= -\left(\frac{1}{a^k}\right)\mathbf{u}_y''^{ikT}\mathbf{T}_i''^k\text{,}\\
        f_y''^{2k} &= \left(\frac{1}{a^k}\right)\mathbf{u}_x''^{ikT}\mathbf{T}_i''^k\text{,}\\
        f_z''^{1k} &= \mathbf{u}_z''^{ikT}\mathbf{F}_i''^k\text{,}\\
        f_x''^{3k} &= \left(\frac{1}{2\,b^k}\right)\mathbf{u}_z''^{ikT}\mathbf{T}_i''^k\text{.}
    \end{align}
    \label{eq_components}
\end{subequations}
The superscripts $1$ and $2$ refer to the components at the two end points of the common joint axis $\mathbf{u}^{\prime\prime}_z$. The distance between these points is the length of the joint $a^k$. The superscript $3$ indicates the components at the end points of the transverse width $b^k$ of the joint geometry along the axis $\mathbf{u}^{\prime\prime}_x$. Naturally, the components $f_z''^{1k}$ and $f_x''^{3k}$ will exist only for revolute and translational joints respectively. 

\begin{enumerate}
    \item For a cylindrical joint, the resultant normal force at both end points of the joint will contribute to the axial friction force. Considering a parabolic force distribution, the effective normal force is given by $F_{n}^k = f^{\prime \prime 1k} +f^{\prime \prime 2k}$ where $f^{\prime \prime lk} = \frac{\pi^3}{24}\sqrt{(f_x^{\prime \prime lk})^2 + (f_y^{\prime \prime lk})^2 + \epsilon^2}$ and $l = 1,\,2$. Additionally, a friction torque will be introduced at each end point; its magnitude is given by $\tau = r_{e}\,\text{cabs}(F_f^{\prime \prime lk})$, where $r_{e}$ is the effective joint radius and $\text{cabs}(x)$ is a continuous approximation to the absolute function defined as $\sqrt{x^2 + \epsilon^2}$, while $\epsilon$ is some small scalar. 
    \item For a revolute joint, the treatment will be similar to a cylindrical joint with the addition of a thrust force in the $z^{\prime\prime}$ direction given by $\mathbf{u}_z''^{ikT}\mathbf{F}_i''^k$. This thrust force will contribute to a frictional torque about an effective torque radius $r_e$.
    \item For a translational joint, the effective normal force is the absolute summation of all the force components. This is given by $F_n^k=\text{cabs}(f_x^{\prime\prime 1k}) + \text{cabs}(f_y^{\prime\prime 1k}) + \text{cabs}(f_x^{\prime\prime 2k}) + \text{cabs}(f_y^{\prime\prime 2k}) + 2\,\text{cabs}(f_x^{\prime\prime 3k})$.
\end{enumerate}
The normal force for any joint type calculated using Equations (\ref{eq_components}a-\ref{eq_components}f) can be plugged in Equation (\ref{eq_S}) to get the scalar frictional force $f_{ij}$ and torque $\tau_{ij}$. Consequently, the generalized friction force vector to be used in the equations of motion can be computed for a pair of bodies $i$ and $j$ as shown in the following equation:
\begin{subequations}
    \begin{align}
    \mathbf{Q}^{Af}_i = \left[\begin{array}{c}
         \mathbf{A}_i\mathbf{C}_i^k\mathbf{u}_z^{\prime\prime ik} f_{ij}\\
          \mathbf{B}^{\text{T}}_i\mathbf{A}_i\mathbf{C}_i^k\mathbf{u}_z^{\prime\prime ik} f_{ij} + 2\mathbf{E}^{\text{T}}_i\mathbf{A}_i\mathbf{C}_i^k\mathbf{u}_z^{\prime\prime ik}\tau_{ij}
    \end{array}\right],\\
        \mathbf{Q}^{Af}_j = \left[\begin{array}{c}
         -\mathbf{A}_i\mathbf{C}_i^k\mathbf{u}_z^{\prime\prime ik} f_{ij}\\
          -\mathbf{B}^{\text{T}}_j\mathbf{A}_i\mathbf{C}_i^k\mathbf{u}_z^{\prime\prime ik} f_{ij} - 2\mathbf{E}^{\text{T}}_j\mathbf{A}_i\mathbf{C}_i^k\mathbf{u}_z^{\prime\prime ik}\tau_{ij}
    \end{array}\right].
\end{align}
\label{eq_gen_fric_force_body_i}
\end{subequations}
These individual generalized friction force vectors can be assembled to give the combined generalized friction force vector of the entire multibody system. Depending upon the assembly of the generalized coordinate vector $\mathbf{q}$, care must be taken to ensure the frictional forces are being appropriately added to the vector of external applied forces and frictional torques to externally applied torques.
\subsection{Equations of motion}
Scleronomic multibody systems in centroidal generalized coordinates are governed by 2\textsuperscript{nd} order index-3 differential-algebraic equations. The index can be reduced by successively differentiating the algebraic constraints to obtain an index-1 augemented form \cite{Serban1997_AugI1DAE} as shown below:
\begin{equation}
    \left[\begin{array}{cc}
        \mathbf{M} \quad& \bm\Phi_{\mathbf{q}}^{\mathrm{T}} \\
         \bm\Phi_{\mathbf{q}} \quad& \mathbf{0}
    \end{array} \right] \left\{ \begin{array}{c}
         \ddot{\mathbf{q}} \\
         \bm\lambda 
    \end{array} \right\} = \left\{ \begin{array}{c}
         \mathbf{Q} + \mathbf{Q}^{Af} \\
         \mathbf{c} 
    \end{array} \right\},
    \label{eq_augmented_index_1}
\end{equation}
where $\mathbf{M}(\mathbf{q},\, \bm\rho)\in \mathbb{R}^{n\times n}$ is the generalized mass matrix, $\bm\Phi(\mathbf{q},\, \bm\rho),\,\bm\lambda \in\mathbb{R}^m$ are the constraints and the associated Lagrange multipliers respectively, $\mathbf{q}\in\mathbb{R}^n$ are the generalized coordinates locating the bodies of the system, $\mathbf{Q}(\mathbf{q},\,\dot{\mathbf{q}},\, \bm\rho)\in\mathbb{R}^n$ and $\mathbf{Q}^{Af}(\mathbf{q},\,\dot{\mathbf{q}},\,\bm\lambda,\, \bm\rho)\in\mathbb{R}^n$ are the generalized external applied and friction forces respectively, $\mathbf{c}(\mathbf{q},\,\dot{\mathbf{q}},\, \bm\rho)\in\mathbb{R}^m$ is an acceleration term equal to  $-(\bm\Phi_\mathbf{q}\dot{\mathbf{q}})_{\mathbf{q}}\dot{\mathbf{q}}$, and $\bm\rho\in\mathbb{R}^p$ is the vector of design and/or control parameters. It is implied in this analysis that the constraint Jacobian $\bm\Phi_\mathbf{q}$ is full rank. For redundant constraints, the reader is referred to Garc{\'i}a de and Guti{\'e}rrez-L{\'o}pez \cite{Garcia2013}. The motion of multibody systems with friction is characterized by sharp changes in acceleration. Moreover, friction is a non-linear function of Lagrange multipliers, making the equations fully-implicit. Additionally, these equations are prone to constraint drift due to enforcement of $\ddot{\bm\Phi}$ instead of $\bm\Phi$. Since these dynamic equations will be differentiated to obtain the TLMs, any errors in the solution of dynamics may further exaggerate the errors in sensitivities. Hence, these equations require some form of implicit integration scheme to contain error propagation with time. Implicit integration schemes help reduce the perturbations introduced in the acceleration field and are more efficient due to the stiff nature of index-1 formulation.\\
We can rewrite (\ref{eq_augmented_index_1}) as an explicit first-order DAE expressed in constant singular mass-matrix form, as shown in the following equations
\begin{subequations}
\begin{align}
\left[\begin{array}{ccc}
    \mathbf{I}\quad & \mathbf{0}\quad & \mathbf{0}\\
         \mathbf{0}\quad &  \mathbf{I}\quad & \mathbf{0}\\
         \mathbf{0}\quad & \mathbf{0}\quad & \mathbf{0}
    \end{array} \right]
 \left\{ \begin{array}{c}
        \dot{\mathbf{q}} \\
         \ddot{\mathbf{q}} \\
         \dot{\bm\lambda} 
    \end{array} \right\} & =\left[\begin{array}{ccc}
        \mathbf{I}\quad & \mathbf{0}\quad & \mathbf{0}\\
        \mathbf{M}\quad & \bm\Phi_{\mathbf{q}}^{\mathrm{T}}\quad & \mathbf{0} \\
         \bm\Phi_{\mathbf{q}} \quad& \mathbf{0}\quad & \mathbf{0}
    \end{array} \right]^{-1} \left\{ \begin{array}{c}
        \dot{\mathbf{q}}\\
         \mathbf{Q} + \mathbf{Q}^{Af} \\
         \mathbf{c} 
    \end{array} \right\} - \left\{ \begin{array}{c}
        \mathbf{0} \\
         \mathbf{0} \\
         \bm\lambda 
    \end{array} \right\} \\
    \implies \Bar{\mathbf{M}}\dot{\mathbf{y}} &= \mathbf{f}(\mathbf{t, y})
\end{align}    
\label{eq_first_order_DAE}
\end{subequations}
where $\mathbf{y} = \left[\mathbf{q}^{\text{T}},\,\mathbf{\dot{q}}^{\text{T}},\,\bm{\lambda}^{\text{T}}\right]^{\text{T}}$. These are linearly implicit differential-algebraic equations \cite{wanner1996solving,steinebach1995order,Steinebach2023,hairer1999} that are solved over a time interval $[t_0,\,t_f]$. Rosenbrock-Wanner (ROW) methods are well known for solving such problems. Another approach for solving these equations involves performing a fixed point iteration every time step \cite{Verulkar2022}. However, fixed point iteration tends to be slow, as it is an additional convergence iteration apart from the implicit integration needed due to solve the acceleration terms. A ROW scheme with stage-number $s$ for the problem is defined by: 
\begin{align}
(\Bar{\mathbf{M}} - h\gamma \mathbf{f}_\mathbf{y}) k_i &= h\mathbf{f}\left(t_0 + \alpha_ih,\,\mathbf{y}_p + \sum_{j=1}^{i-1}\alpha_{ij}k_j\right) + h \mathbf{f}_\mathbf{y}\sum_{j=1}^{i-1}\gamma_{ij}k_{j} + h^2\gamma_{i}\mathbf{f}_t\\
\mathbf{y}_{p+1} &= \mathbf{y}_p + \sum_{i=1}^{s}b_ik_i, \quad i=1,...,s\quad p=0,1,..., \\
\text{where } \mathbf{f}_\mathbf{y} &= \frac{\partial \mathbf{f}}{\partial \mathbf{y}}(t_p, \mathbf{y}_p),\\
\mathbf{f}_t &= \frac{\partial \mathbf{f}}{\partial t}(t_p, \mathbf{y}_p),
\end{align}
where $h$ is the time step for integrator step $p$. The coefficients of the method are $\gamma$, $\alpha_{ij}$, $\gamma_{ij}$, and $b_i$ define the weights. Moreover, it holds $\alpha_i = \sum_{j = 1}^{i-1} \alpha_{ij}$ and $\gamma_i = \gamma + \sum_{j=1}^{i-1}\gamma_{ij}$. It is worth noting that this scheme works well only for index-1 systems, since it guarantees the regularity of the matrix $(\Bar{\mathbf{M}} - h\gamma \mathbf{f}_\mathbf{y})$. In addition to the ROW methods described before, a fully-implicit Runge-Kutta method (\texttt{Radau}) \cite{hairer2010solving} is also suitable for solving Equation (\ref{eq_first_order_DAE}). However, multistep BDF methods like \texttt{CVODE}/\texttt{FBDF} \cite{Hindmarsh2005} and \texttt{DASSL} \cite{osti_5882821} were found to be comparatively slow for these problems and also led to failed integration due to high integration errors for some problems. Also, it is possible to solve the equations as fully implicit DAEs in residual form; however, obtaining consistent initial conditions is difficult and may lead to integration failure.

The positions $\mathbf{q}_0$ at the initial time $t_0$ can be obtained by introducing a set of temporary constraints $\bm\Psi\in\mathbb{R}^{n-m}$ to complete the non-linear system of constraint equations and make the constraint Jacobian square. Exact generalized coordinates $\mathbf{q}$ can be converged from a given initial estimate using the following Newton-Raphson iterative scheme:
\begin{subequations}
    \begin{align}
    \Bar{\bm\Phi}=\left\{\begin{array}{c}
    \bm{\Phi}\\
    \bm{\Psi}
    \end{array}\right\}_{t_0} &= \mathbf{0},\\
\left[\begin{array}{cc}
       \bm{\Phi}_{\mathbf{q}}  & \mathbf{0} \\
        \mathbf{0} & \bm{\Psi}_{\mathbf{q}}
    \end{array}\right]_{t_0}\,\Delta{\mathbf{q}}^i &= -\left\{\begin{array}{c}
    \bm{\Phi}\\
    \bm{\Psi}
    \end{array}\right\}_{t_0},\\
    \mathbf{q}^{i+1} &= \mathbf{q}^i + \Delta{\mathbf{q}}^i.
    \label{eq_newton_iter}
\end{align}
\end{subequations}
For spatial systems, obtaining a reasonably good estimate of $\mathbf{q}$ is a challenge and typically requires a CAD model. A trust-region algorithm \cite{NoceWrig06,osti_6997568,RePEc:mtp:titles:0262633094} from the \texttt{NLSolve.jl} package was used to converge all initial estimates to machine precision.

A trivial solution for the initial velocity vector is $\dot{\mathbf{q}} = \mathbf{0}$ since it will always satisfy the velocity constraints $\dot{\bm\Phi} = \bm\Phi_{\mathbf{q}}\dot{\mathbf{q}} = \mathbf{0}$ which implies the system is at rest. For non-zero initial velocities, the solution has to be a linear combination of the null space vectors of $\bm\Phi_\mathbf{q}$ at the initial time such that the velocities of the ($n-m$) chosen independent generalized coordinates are satisfied. Alternatively, the velocity constraints $\dot{\bm\Phi}$ can be completed to obtain the generalized velocities \cite{Haug2021}. Since non-zero initial velocities lead to non-negligible frictional forces, it is relatively difficult to obtain initial Lagrange multipliers. In this analysis, we obtain the estimate for initial Lagrange multipliers by ignoring $\mathbf{Q}^{Af}$, temporarily making the equations of motion explicit. Thereafter, an exact solution is obtained through a non-linear solution of the residual shown below
\begin{align}
    \mathbf{R} = \left\{\begin{array}{c}
        \ddot{\mathbf{q}}   \\
          \bm\lambda
    \end{array}\right\} - \left[\begin{array}{cc}
        \mathbf{M} \quad& \bm\Phi_{\mathbf{q}}^{\mathrm{T}} \\
         \bm\Phi_{\mathbf{q}} \quad& \mathbf{0}
    \end{array} \right]^{-1}\left[ \begin{array}{c}
         \mathbf{Q} + \mathbf{Q}^{Af} \\
         \mathbf{c} 
    \end{array} \right] \to \mathbf{0}
\label{eq_fixed_point}
\end{align}
%
\section{Gradient-based optimization}
\label{sec:optimization}
Dynamic optimization aims to find the optimal design or control parameters while minimizing or maximizing a specified objective function, subject to constraints imposed by the system dynamics and parameter/state bounds. This can be mathematically expressed as an initial value bound-constrained optimization problem as shown in the following equations:
\begin{subequations}
\begin{align}
            \min_{\bm\rho}\quad &\bm\psi(\mathbf{y}, \bm\rho),\\
    \text{such that}\quad \Bar{\mathbf{M}}\dot{\mathbf{y}} &= \mathbf{f}\left(\mathbf{y}, \bm\rho, t\right),\\
    \quad\underline{\bm\rho}\, \leq \, &\bm\rho\, \leq\,  \overline{\bm\rho},\\
        \text{and}\quad \mathbf{y} \big\rvert_{t_0} &= \mathbf{y}_0,
\end{align}
\label{eq_gen_optimization}
\end{subequations}
where $\underline{\bm\rho}$,  $\overline{\bm\rho}$  $\in \mathbb{R}^p$ represent the lower and upper bound vectors respectively. We will mainly deal with the objective function of the form:
\begin{align}
    \bm\psi = \mathbf{w}\left(\mathbf{y}, \bm{\rho} \right)\Big\rvert_{t_f} + \int_{t_o}^{t_f}\mathbf{g}\left(\mathbf{y}, \bm{\rho}\right)\text{d}t.
    \label{eq_obj_fcn}
\end{align}  
In this context, $\bm\psi\in\mathbb{R}^o$ represents a vector containing '$o$' objective functions. Additionally, $\mathbf{w}\in\mathbb{R}^o$ corresponds to the pointwise term at the final time, while $\mathbf{g}\in\mathbb{R}^o$ denotes the integrand associated with each of the objective functions.
\begin{align}
    \min_{\bm\rho,\, \mathbf{u}}\int_{t_0}^{t_f} \left(\mathbf{y - y}_{\text{ref}}\right)^{\text{T}} \mathbf{Q} \left(\mathbf{y - y}_{\text{ref}}\right) + \mathbf{u}^{\text{T}} \mathbf{R}\mathbf{u}\,\text{d}t.
    \label{eq:LQR_obj_func}
\end{align}
The quadratic cost function shown in Equation (\ref{eq:LQR_obj_func}) is a widely applicable specialization of the objective function (\ref{eq_obj_fcn}) for control case studies. Here, the control $\mathbf{u} =  \mathbf{h}(\bm\rho, t)$ contained in a ball of admissible control set $\bm\Omega$ is required to follow a trajectory $\mathbf{y}_{\text{ref}}$ and $\mathbf{Q}$ and $\mathbf{R}$ are positive semi-definite penalty matrices for state error and control, respectively. The motivation behind this choice of objective function is that most optimization problems for dynamic systems can be formulated either as trajectory tracking problems or regulating problems. Time optimal problem formulations of the form
\begin{align}
    \min_{\bm\rho}\quad &t_f \\
    \text{such that}\quad \mathbf{y} \big\rvert_{t_f} &= \mathbf{y}_f\\
    \text{and}\quad \mathbf{y} \big\rvert_{t_0} &= \mathbf{y}_0
\end{align}
are out of scope for this research. The gradient of the objective function (\ref{eq_obj_fcn}) is given by:
\begin{align}
    \begin{split}
&\nabla_{\bm{\rho}}\bm\psi^\text{T}=\left(\mathbf{w}_{\mathbf{q}}\mathbf{q}^{\prime}+\mathbf{w}_{\dot{\mathbf{q}}}\dot{\mathbf{q}}^{\prime}+\mathbf{w}_{\bm{\lambda}}\bm{\lambda^{\prime}}+\mathbf{w}_{\bm{\rho}}\right)\Big\rvert_{t_f} \\
      &\qquad+\int_{t_o}^{t_f} \left(\mathbf{g}_{\mathbf{q}}\mathbf{q}^{\prime}+\mathbf{g}_{\dot{\mathbf{q}}}\dot{\mathbf{q}}^{\prime}+\mathbf{g}_{\bm{\lambda}}\bm\lambda^{\prime}+\mathbf{g}_{\bm{\rho}}\right) \text{d}t,
      \end{split}
      \label{eq_grad_obj_fcn}
\end{align}
where the terms $\mathbf{q}^{\prime},\,\dot{\mathbf{q}}^{\prime}\in\mathbb{R}^{n\times p}$, and $\bm{\lambda}^{\prime}\in\mathbb{R}^{m\times p}$ are the respective state sensitivities (total derivatives) with respect to the parameters $\bm\rho$. To obtain these sensitivities of the states, we differentiate Equation (\ref{eq_augmented_index_1}) with respect to the parameters $\bm\rho$, as shown below:
\begin{subequations}
\begin{align}
    \frac{\text{d}\mathbf{M}(\mathbf{q})\ddot{\mathbf{q}}}{\text{d}\bm\rho}+\frac{\text{d}\bm{\Phi}^\text{T}_{\mathbf{q}}(\mathbf{q})\bm{\lambda}}{\text{d}\bm\rho}&=\frac{\text{d}(\mathbf{Q}^{Af}(\mathbf{q}, \dot{\mathbf{q}}, \bm{\lambda}) +\mathbf{Q}(\mathbf{q}, \dot{\mathbf{q}}, t))}{\text{d}\bm\rho},
        \\
        \frac{\text{d}\ddot{\bm{\Phi}}}{\text{d}\bm\rho} &= \bm{0}.
\end{align}
\label{eq_deriv_EOM}
\end{subequations}
If we consider that all terms are dependent on $\bm\rho$, the total derivatives can be expanded, resulting in the final TLMs as shown below:
%
%
\begin{subequations}
\begin{align}
       \mathbf{M}\ddot{\mathbf{q}}^{\prime} + \Bar{\mathbf{C}} \dot{\mathbf{q}}^{\prime} + \Bar{\mathbf{K}} \mathbf{q}^{\prime} + \Bar{\mathbf{L}} \bm\lambda^{\prime} &= \Bar{\mathbf{Q}},\quad\text{and}\\
       \bm\Phi_\mathbf{q}\ddot{\mathbf{q}}^{\prime}-\mathbf{c}_{\dot{\mathbf{q}}}\dot{\mathbf{q}}^{\prime} + (\bm\Phi_{\mathbf{q}\mathbf{q}}\ddot{\mathbf{q}} - \mathbf{c}_\mathbf{q})\mathbf{q}^\prime &= \mathbf{c}_{\bm\rho} - \bm\Phi_{\mathbf{q}\bm\rho}\ddot{\mathbf{q}},
\end{align}
\label{eq_sensitivities}
\end{subequations}
where,
\begin{subequations}
    \begin{align}
       \Bar{\mathbf{C}} &= -\mathbf{Q}_{\dot{\mathbf{q}}} -{\mathbf{Q}^{Af}_{\dot{\mathbf{q}}}},\\
       \Bar{\mathbf{K}} &=\mathbf{M}_\mathbf{q}\ddot{\mathbf{q}} +{\bm\Phi^\text{T}_{\mathbf{q}\mathbf{q}}}\bm\lambda - \mathbf{Q}_\mathbf{q} -{\mathbf{Q}^{Af}_\mathbf{q}},\\
       \Bar{\mathbf{L}} &= \bm\Phi^\text{T}_\mathbf{q}-{\mathbf{Q}^{Af}_{\bm\lambda}},\\
       \Bar{\mathbf{Q}} &=\mathbf{Q}_{\bm\rho}+\mathbf{Q}^{Af}_{\bm\rho}- \mathbf{M}_{\bm\rho}\ddot{\mathbf{q}} - {\bm\Phi^\text{T}_{\mathbf{q}\bm\rho}}\bm\lambda,\\
       \mathbf{c}_\mathbf{q} &= -\dot{\bm\Phi}_\mathbf{qq}\dot{\mathbf{q}},\\
       \mathbf{c}_{\dot{\mathbf{q}}} &= -\dot{\bm\Phi}_\mathbf{q\dot{q}}\dot{\mathbf{q}} - \dot{\bm\Phi}_\mathbf{q},\\
       \mathbf{c}_{\bm\rho} &= -\dot{\bm\Phi}_{\mathbf{q}\bm\rho}\dot{\mathbf{q}}.
    \end{align}
    \label{eq_sensitivities_terms}
\end{subequations}
In Equation (\ref{eq_sensitivities_terms}), frictional force vector $\mathbf{Q}^{Af}$ and the corresponding Jacobians $\mathbf{Q}^{Af}_{\mathbf{q}}$, $\mathbf{Q}^{Af}_{\dot{\mathbf{q}}}$, $\mathbf{Q}^{Af}_{\bm\lambda}$, and $\mathbf{Q}^{Af}_{\bm\rho}$ are all dependent on Lagrange multipliers. It is important to note that Equations (\ref{eq_sensitivities_terms}) contain several terms such as $\mathbf{M_q}\ddot{\mathbf{q}}$, $\bm\Phi_{\mathbf{qq}}^\text{T}\bm{\lambda}$, and so on that are tensor-vector products. These can be computed for any matrix $\mathbf{A} \in \mathbb{R}^{q\times r}$ and any pair of vectors $\mathbf{b}\in\mathbb{R}^r$ and $\mathbf{x}\in\mathbb{R}^s$ as shown below:
\begin{subequations}
   \begin{align}
    \mathbf{A}_{\mathbf{x}} = \left[\frac{\partial \mathbf{A}}{\partial x_1},\quad\cdots,\quad\frac{\partial \mathbf{A}}{\partial x_i},\quad\cdots,\quad\frac{\partial \mathbf{A}}{\partial x_s}\right]\in\mathbb{R}^{q\times r\times s}\\
    \mathbf{A}_{\mathbf{x}}\mathbf{b} = \left[\frac{\partial \mathbf{A}}{\partial x_1}\mathbf{b},\quad\cdots,\quad\frac{\partial \mathbf{A}}{\partial x_i}\mathbf{b},\quad\cdots,\quad\frac{\partial \mathbf{A}}{\partial x_s}\mathbf{b}\right]\in\mathbb{R}^{q\times s}
\end{align} 
\end{subequations}
However, as differentiation is a linear operation, tensor algebra can be easily avoided by premultiplying the respective matrices with the associated vector before differentiating $\mathbf{A}_{\mathbf{x}}\mathbf{b} = \left(\mathbf{A}\mathbf{b}\right)_{\mathbf{x}}$.

To solve Equation (\ref{eq_sensitivities}), a set of $2np$ initial conditions is necessary, represented by position sensitivities $\mathbf{q}^{\prime}\big\vert_{t_0} = {\mathbf{q}}^{\prime}_0$ and velocity sensitivities $\dot{\mathbf{q}}^{\prime}\big\vert_{t_0} = \dot{\mathbf{q}}^{\prime}_0$. As these initial sensitivities must adhere to the sensitivity constraints at the initial time, they can be determined by solving the following equation:
\begin{subequations}
\begin{align}
    \frac{\text{d}\Bar{\bm\Phi}}{\text{d}\bm\rho}\Bigg\vert_{t_0} = \mathbf{0} &\implies \Bar{\bm\Phi}_{\mathbf{q}}\big\vert_{t_0}\,{\mathbf{q}}^{\prime}_0 = -\Bar{\bm\Phi}_{\bm\rho}\big\vert_{t_0},\\
    \frac{\text{d}\dot{\Bar{\bm\Phi}}}{\text{d}\bm\rho}\Bigg\vert_{t_0} = \bm{0} &\implies \Bar{\bm\Phi}_{\mathbf{q}}\big\vert_{t_0}\,{\dot{\mathbf{q}}}^{\prime}_0 = -\left(\Bar{\bm\Phi}_{\mathbf{qq}}\mathbf{q}^{\prime}_0+\Bar{\bm\Phi}_{\mathbf{q}\bm\rho}\right)\dot{\mathbf{q}}\big\vert_{t_0}.
\end{align}
\label{eq_initial_sen}
\end{subequations}
It can be observed from Equation (\ref{eq_initial_sen}b) that if the system starts at rest then the velocity sensitivities are also zero. Rearranging Equation (\ref{eq_sensitivities}) in an augmented matrix form we have:
\begin{align}
    \left[\begin{array}{cc}
    \mathbf{M} \quad& \Bar{\mathbf{L}}\\
    \bm\Phi_{\mathbf{q}} \quad& \mathbf{0}
    \end{array}\right] \left\{\begin{array}{c}
    {\ddot{\mathbf{q}}}^{\prime} \\
    {\bm{\lambda}}^{\prime}
    \end{array}\right\}= \left[\begin{array}{c}
        \mathbf{A}\\
        \mathbf{B}
    \end{array}\right],
    \label{eq_matrix_sen}
\end{align}
%
%
%
where,
\begin{subequations}
    \begin{align}
        \mathbf{A} &= \Bar{\mathbf{Q}} - \Bar{\mathbf{C}}\dot{\mathbf{q}}^{\prime} - \Bar{\mathbf{K}}\mathbf{q}^{\prime},\\
        \mathbf{B} &= \mathbf{c}_{\bm\rho} - \bm\Phi_{\mathbf{q}\bm\rho}\ddot{\mathbf{q}} + \mathbf{c}_{\dot{\mathbf{q}}}\dot{\mathbf{q}}^{\prime}- (\bm\Phi_{\mathbf{q}\mathbf{q}}\ddot{\mathbf{q}} - \mathbf{c}_\mathbf{q})\mathbf{q}^{\prime}.
    \end{align}
\end{subequations}
Equation (\ref{eq_matrix_sen}) is an index-1 DAE since the algebraic variable $\bm\lambda^{\prime}$ does not appear as a differential term anywhere in the sensitivity mass-matrix on the left or the matrices $\mathbf{A}$ and $\mathbf{B}$ on the right. It can be solved for $\ddot{\mathbf{q}}^{\prime}$ and $\bm{\lambda}^{\prime}$ through matrix inversion and integrated to obtain the sensitivities $\dot{\mathbf{q}}^{\prime}$ and $\mathbf{q}^{\prime}$. All sensitivities can be simultaneously computed using a solver for differential equations in matrix form or by reshaping the associated matrices at every iteration to make the system compatible with standard vector form ODE solvers. The sensitivities for multibody systems using the Brown and McPhee friction model tend to exhibit abrupt jumps, akin to those seen in the hybrid dynamic systems \cite{Verulkar2022}. Ideally, a stiffness-aware integrator with automatic switching, like \texttt{LSODA} \cite{hindmarsh1983odepack,radhakrishnan1993description,doi:10.1137/0904010}, should be used for obtaining the sensitivities.

The gradient of the objective function can be calculated for all the parameters once the sensitivities have been obtained and the optimization iterations can be started. The study used quasi-Newton methods, predominantly L-BFGS, which are suitable for for bound-constrained optimization problems solved with this methodology. They require only the first-order gradient of the objective function which is essential for multibody optimizations, since computing the Hessian matrix would be mathematically and computationally infeasible. The algorithm is described briefly for readers convenience in the context of multibody optimization. For more details the reader is referred to \cite{LBFGS_bound,Rothwell2017}. This algorithm approximates the cost function through the following quadratic function:
\begin{align}
    \psi(\bm\rho) \approx \psi(\bm\rho_k) + \nabla \psi (\bm\rho_k)^T (\bm\rho - \bm\rho_{k}) + \frac{1}{2}(\bm\rho - \bm\rho_{k})^T \mathbf{B}_k(\bm\rho - \bm\rho_{k}),
\end{align}
where $\mathbf{B}_k$ is the limited-memory approximation for the Hessian matrix at iteration $k$. A piecewise linear path is assumed for the design parameters:
\begin{subequations}
    \begin{align}
    \bm\rho_{k+1} &= \mathbf{P}(\bm\rho_{k} - t \nabla \psi (\bm\rho_k),\, \underline{\bm\rho},\, \overline{\bm\rho}),\\
    \mathrm{where} \, \mathbf{P}(\bm\rho, \underline{\bm\rho}, \overline{\bm\rho}) &= \text{max}(\underline{\bm\rho}, \text{min}(\bm\rho,\,\overline{\bm\rho})).
\end{align}
\label{eq:Cauchy}
\end{subequations}
Equations (\ref{eq:Cauchy}) determine the Cauchy point $\bm\rho^{c}$, which is the first local minimum of $\psi(\bm\rho)$. The variables whose Cauchy point is at the lower or upper bound are held fixed. These comprise the active-set $\mathcal{A}(\bm\rho^{c})$. The following quadratic problem is considered over the subspace of free variables, to calculate an approximate solution $\bm\rho^{*}_{k+1}$:
\begin{subequations}
  \begin{align}
    \min_{\rho} \{\psi(\bm\rho^{*}) : \rho_i = \rho_i^{c},\, \forall\, i \in \mathcal{A}(\rho^{c})\},\\
    \text{such that} \quad \underline{\rho}_i < \rho_i < \overline{\rho}_i,\, \forall\, i \notin \mathcal{A}(\rho^{c}).
\end{align} 
\label{eq:approx_iter}
\end{subequations}
Equation (\ref{eq:approx_iter}) can be solved in two ways. First, the bounds on the free variables can be ignored for optimization and the solution can be obtained by direct or iterative methods. Then, the free variable path can be truncated toward the solution so as to satisfy the bounds. Another approach is to handle the active bounds by Lagrange multipliers. Once an approximate solution $\bm\rho^{*}_{k+1}$ is obtained, the next iterate  $\bm\rho_{k+1}$ can be obtained by backtracking linear search along $\mathbf{d}_k = \bm\rho^{*}_{k+1} - \bm\rho_k$, that ensures:
\begin{align}
    \psi(\bm\rho_{k+1}) \leq \psi(\bm\rho_k) + \alpha_k\nabla \psi (\bm\rho_k)^{\mathrm{T}}\mathbf{d}_k,
\end{align}
where $\alpha_k$ is the optimal step size. This process is repeated until convergence by evaluating the gradient at $\bm\rho_{k+1}$ and computing a new limited-memory Hessian approximation $\mathbf{B}_{k+1}$. The optimization package \texttt{Optim.jl} and MATLAB \texttt{fminunc} were used to obtain the results of the case studies in this research.
\section{Case studies}
\label{sec:case_studies}
Multibody optimizations tend to be highly non-convex due to the large rotations involved in the dynamics. Hence, even though a convex objective function is chosen, the non-linear dynamics makes the solution heavily dependent on the chosen initial condition. This is usually not a problem though since gradient-based methods converge fast, allowing the user to experiment with different initial parameter estimates. This section covers three optimization case studies. The first two examples will demonstrate the applicability of this methodology to pure control and pure design problems respectively. Finally, an example of co-design is included to showcase the advantages of the proposed methodology over the classical optimization approach.
\subsection{Inverted spatial pendulum}
The inverted pendulum is a classic problem in dynamics and control theory, frequently used as a benchmark for testing control strategies. This system is inherently unstable and requires a feedback control loop for stabilization. Interestingly, many real-world systems behave like an inverted pendulum. All bipedal and humanoid robot motion, as well as the motion of bicycles and motorcycles, is similar to that of an inverted pendulum. In this study, the spatial variant of the inverted pendulum is used with a 2-dimensional PID controller for stabilization. This additional degree of freedom adds substantial complexity in terms of modeling and control computation. The total number of degrees of freedom (DOFs) for such a system is 5 (2 DOFs for the cart translation along the ground XY plane and 3 rotational DOFs for the pendulum). Friction is non-negligible at the interface between the the cart and the ground. This is expected as the weight of the mechanism leads to a high normal reaction at this interface. Friction within the system can be represented through various modeling approaches. The case study is divided into two segments. In the initial part, both the simulation and optimization models incorporate the Brown McPhee friction model. In the subsequent section, focusing on the importance of friction modeling, the optimization model employs the Brown McPhee model, while the simulation uses the Gonthier friction model.

The goal of this study is to use the methodology presented in this paper for computing the optimal gains for the PID controller i.e., $\bm\rho = [K_p,\, K_i,\, K_d$]. Manually tuning a PID controller is difficult, especially for non-linear systems. The methodology described in this paper can be used to convert this optimal feedback control problem into a parameter optimization problem for multibody systems. The optimization is set up to minimize the error in X and Y positions of the cart center $(x_1,\, y_1)$ and the pendulum center $(x_2,\, y_2)$. Thus, we obtain:
\begin{align}
   e_x = x_2-x_1, &\quad \text{and}\quad e_y = y_2 - y_1.
\end{align}
This study has been implemented in MATLAB\textsuperscript{{}\textregistered} using Symbolic Math Toolbox for evaluating Jacobians with respect to design parameters. The objective function chosen for this study is quadratic, with high penalty on the error in states and a comparatively low penalty on the control. This also has a scaling effect since the control magnitude is much higher than that of the error. The final objective function used for this case study is:
\begin{align}
    \psi = (10^5 e_x^2 + 10^5 e_y ^2) \Big\rvert_{t_f}+\int_{t_0}^{t_f} 10^5 e_x^2 + 10^5 e_y ^2 + u_x^2 + u_y^2 \mathrm{d}t,
        \label{eq:inverted_pendulum_obj}
\end{align}
where
\begin{align}
    u_x(t) &= K_p e_x + K_i\int_{t_0}^{t}  e_x d\mathrm{t} + K_d \frac{\mathrm{d}e_x}{\mathrm{d}t}, \quad \text{and}\\
    u_y(t) &= K_p e_y +  K_i\int_{t_0}^{t} e_y d\mathrm{t} + K_d \frac{\mathrm{d}e_y}{\mathrm{d}t}.
\end{align}
Hence, we are penalizing both, the continuous error, as well as the error at the final time. For gradient-based methods to be effective, it is crucial for users to provide a good initial estimate for the parameters. The chosen cost function has a local minimum if the pendulum hangs vertically below the plate but this solution is undesirable. To steer the optimization in the desired direction, the initial control should achieve stabilization to some extent. Typically, control saturation constraints are imposed for such optimization studies to consider the limits of the actuation system. However, since this paper only deals with bound constraint problems, bounding the gains instead of the computed control is not appropriate. Hence, an unconstrained optimization is performed and the control saturation limits are imposed indirectly by controlling the state and control penalties in the objective integrand. This approach is suitable for such studies since the time taken for optimization is small and multiple optimization iterations can be easily performed. Table \ref{tab:inverted_pendulum_eqns} details the equations used for this study.
\begin{table}[htbp]
    \centering
    \caption{Inverted pendulum: number of equations}
    \begin{tabular}{@{}lc@{}}
    \toprule
    \textbf{Component} & \textbf{Value} \\
    \midrule
    Number of bodies & $2$ \\
    States per body & $7$ \\
    Total differentiable variables for dynamics & $2 \times 7 = 14$ \\
    First-order equations of motion & $2 \times 14 = 28$ \\
    Degrees of freedom & $5$ \\
    Lagrange multipliers/Constraints & $14 - 5 = 9$ \\
    Total dynamic equations & $28 + 9 = 37$ \\
    Number of free-variables (parameters) & $3$\\
    Total number of sensitivities & $3\times 37 = 111$\\
    Total differential-algebraic equations & $37+111=148$\\
    Total objective function(s) & $1$\\
    Total objective function gradients & $3$\\
    \bottomrule
    \end{tabular}
    \label{tab:inverted_pendulum_eqns}
\end{table}

As it can be observed in Table \ref{tab:inverted_pendulum_eqns}, a system of 148 differential-algebraic equations must to be solved for every optimization iteration along with the additional integrations required to calculate the objective function and its gradient. This is for a relatively simple control case study. Since optimization studies can take several iterations to converge, the computational cost adds up. This highlights the importance of using efficient computation techniques, like sparse-matrix algebra, iterative solvers, SIMD/parallelism, and efficient memory allocation in functions, in order to speed up the process.

\begin{figure}[htbp]
     \centering
     \begin{subfigure}[b]{\textwidth}
         \centering
         \includegraphics[width=\textwidth]{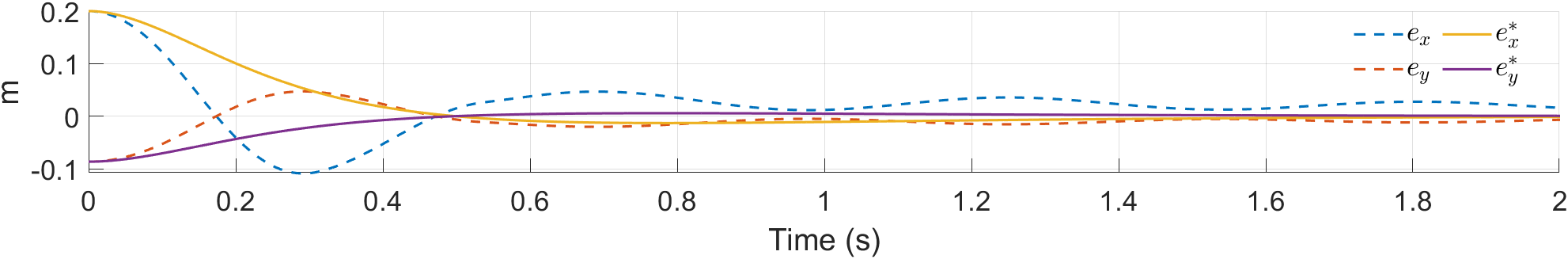}
         \caption{Position error in rod and plate center of mass}
         \label{fig:inv_pend_pos_error}
     \end{subfigure}
     \hfill
     \begin{subfigure}[b]{\textwidth}
         \centering
         \includegraphics[width=\textwidth]{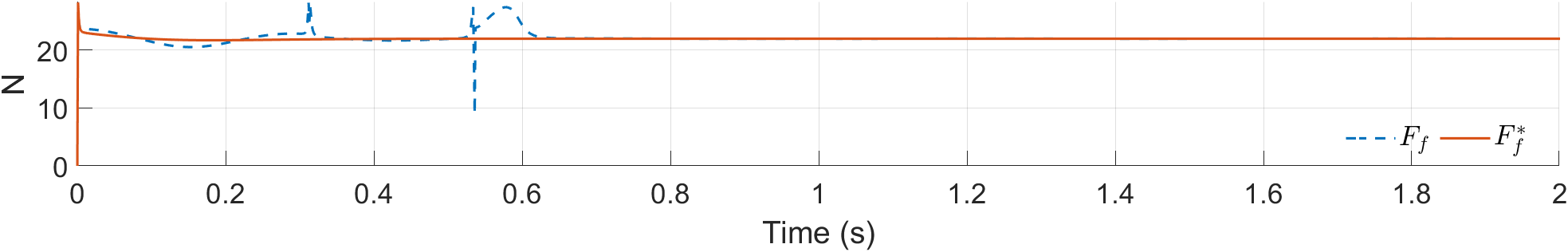}
         \caption{Friction forces at plate-ground interface}
         \label{fig:inv_pend_fric}
     \end{subfigure}
     \hfill
     \begin{subfigure}[b]{\textwidth}
         \centering
         \includegraphics[width=\textwidth]{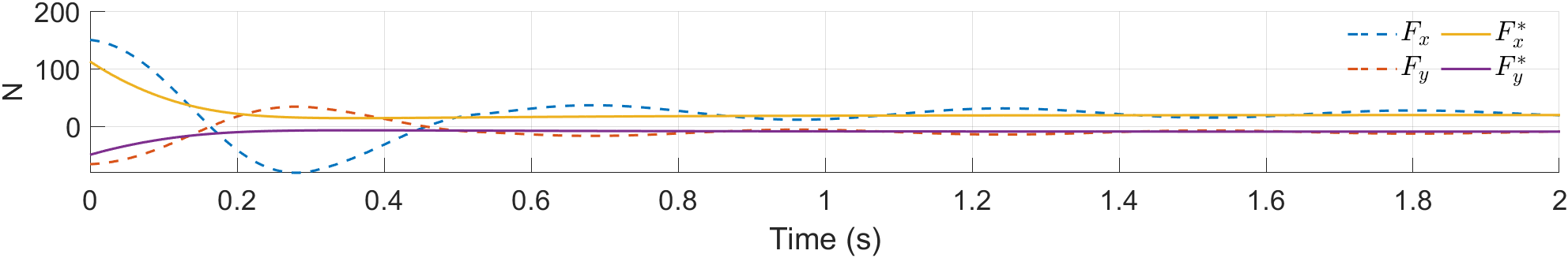}
         \caption{PID control forces acting on the plate}
         \label{fig:inv_pend_cont}
     \end{subfigure}
        \caption{Inverted pendulum dynamics: initial vs. optimal controller parameters}
        \label{fig:inverted_pendulum_dynamics}
\end{figure}

Figure \ref{fig:inverted_pendulum_dynamics} shows the dynamic response of the system before and after optimization. The PID gain initial estimates were [750, 200, 10], which converged post-optimization to [563.11, 626.57, 90.68]. Figure \ref{fig:inverted_pendulum_dynamics}a shows the error dynamics for the system with the initial control versus the optimized control. As it can be seen, the response is significantly better post-optimization. Figure \ref{fig:inverted_pendulum_dynamics}b shows the friction at the plate-ground interface. The pre-optimization friction switched from dynamic friction to stiction and back at 0.3 sec and 0.55 sec notably. This is a consequence of aggressive control overcompensating for the stabilization error. The cart keeps changing its direction of motion causing the friction to behave in a way that opposes this change. The friction force is constant for the optimized control as it provides the optimal compensation to achieve stabilization. Figure \ref{fig:inverted_pendulum_dynamics}c shows the control forces before and after optimization. The maximum control effort and energy used post optimization is much lower, however the control stabilizes the pole in less than 1/5\textsuperscript{th} of the time taken by the unoptimized controller. 

\begin{figure}[htbp]
        \centering
        \subgraphics{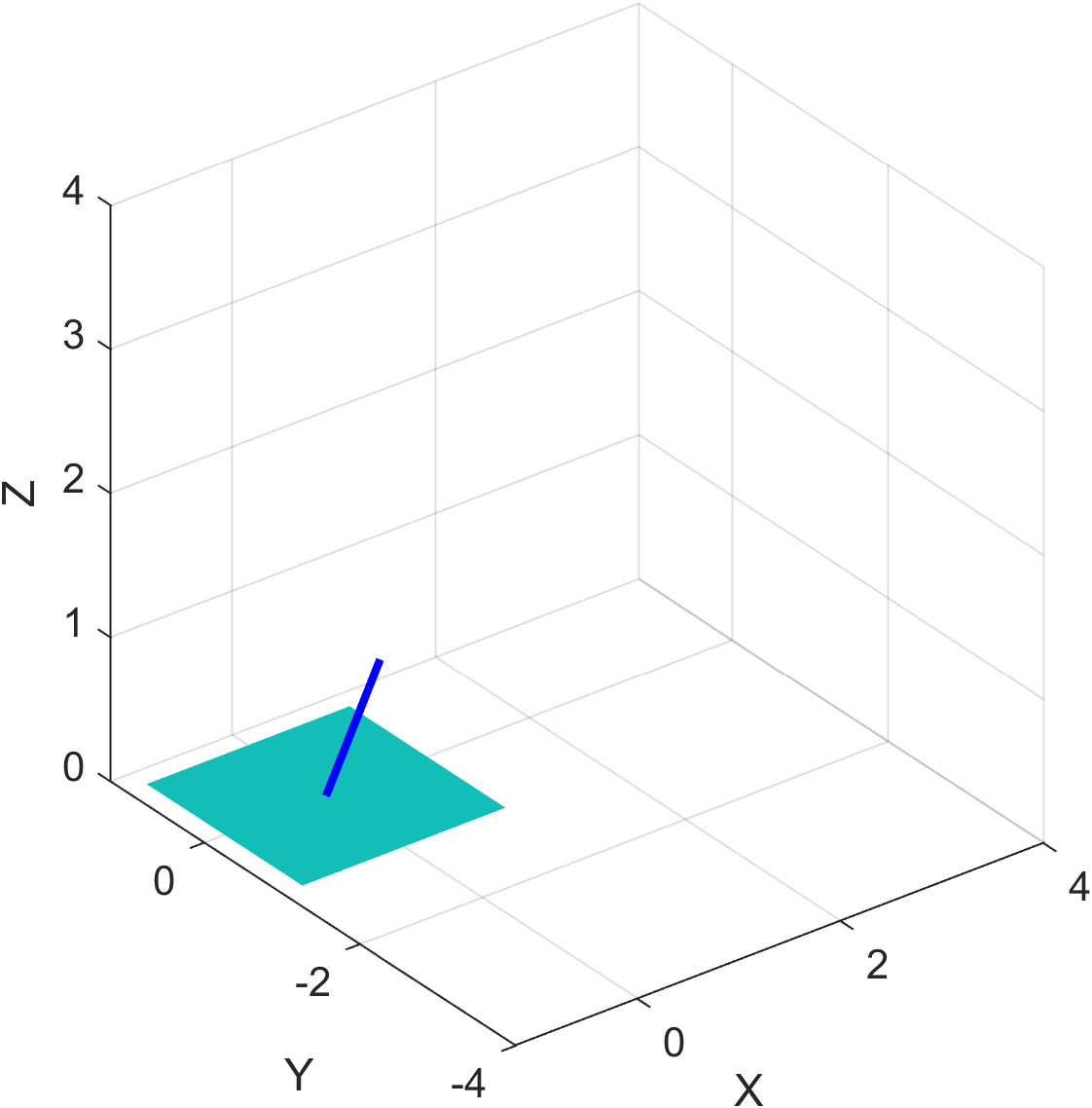}{t = 0}
        \subgraphics{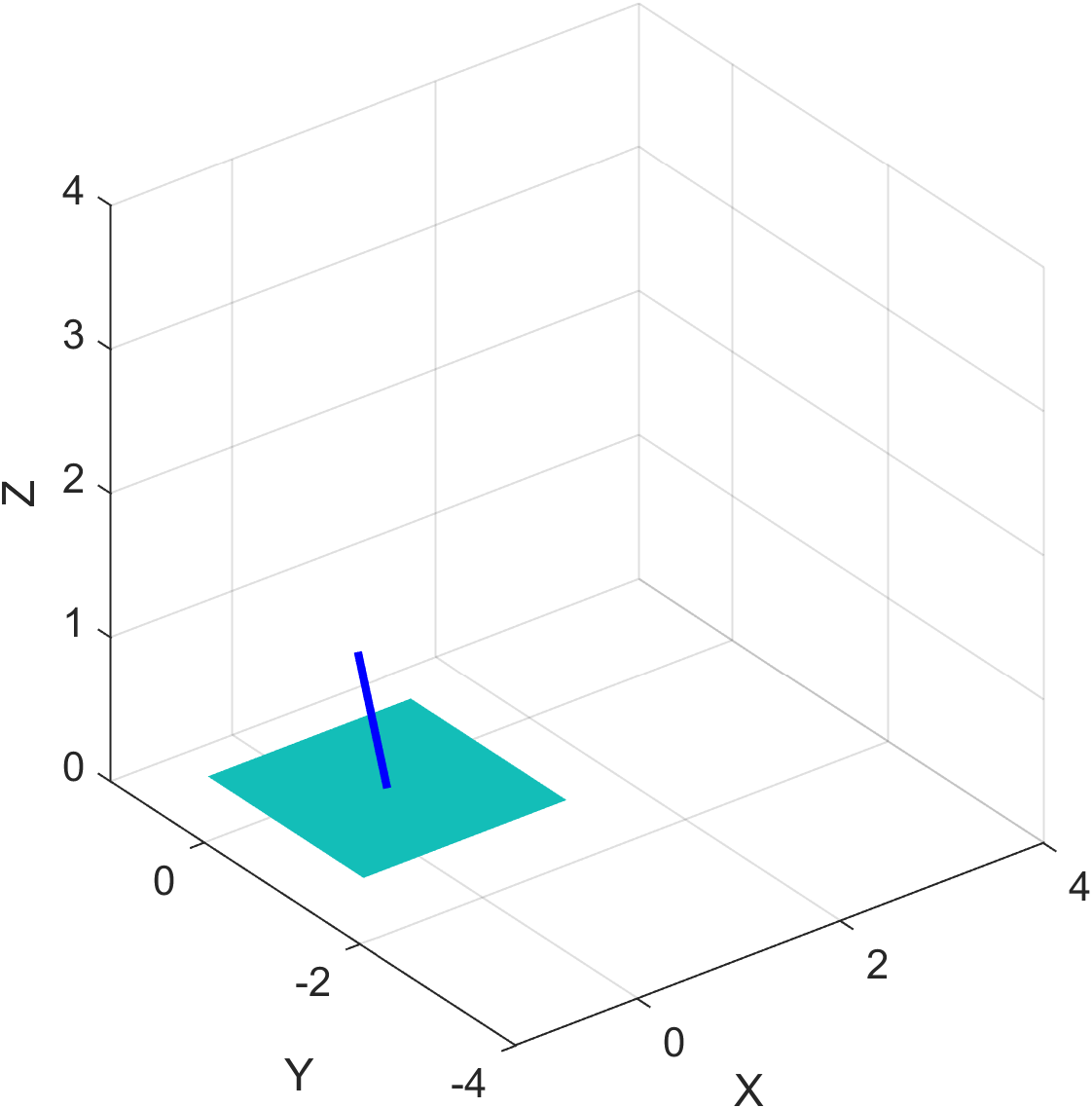}{t = 300 ms}
        \subgraphics{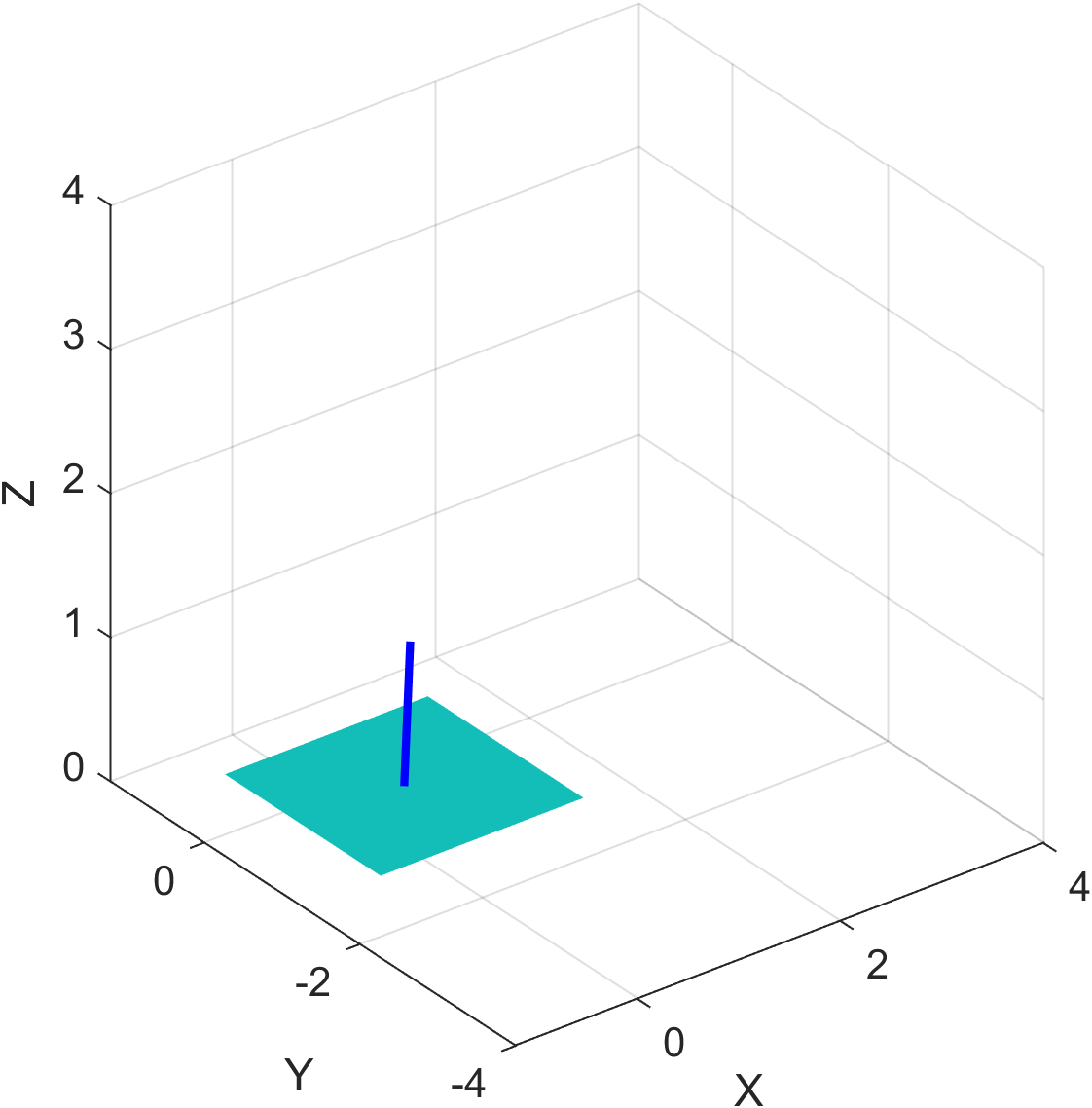}{t = 1100 ms} 
         \caption{Inverted pendulum stabilization : initial PID control parameters}
         \label{fig:inverted_pendulum_anim}
\end{figure}

\begin{figure}[htbp]
        \centering
        \subgraphics{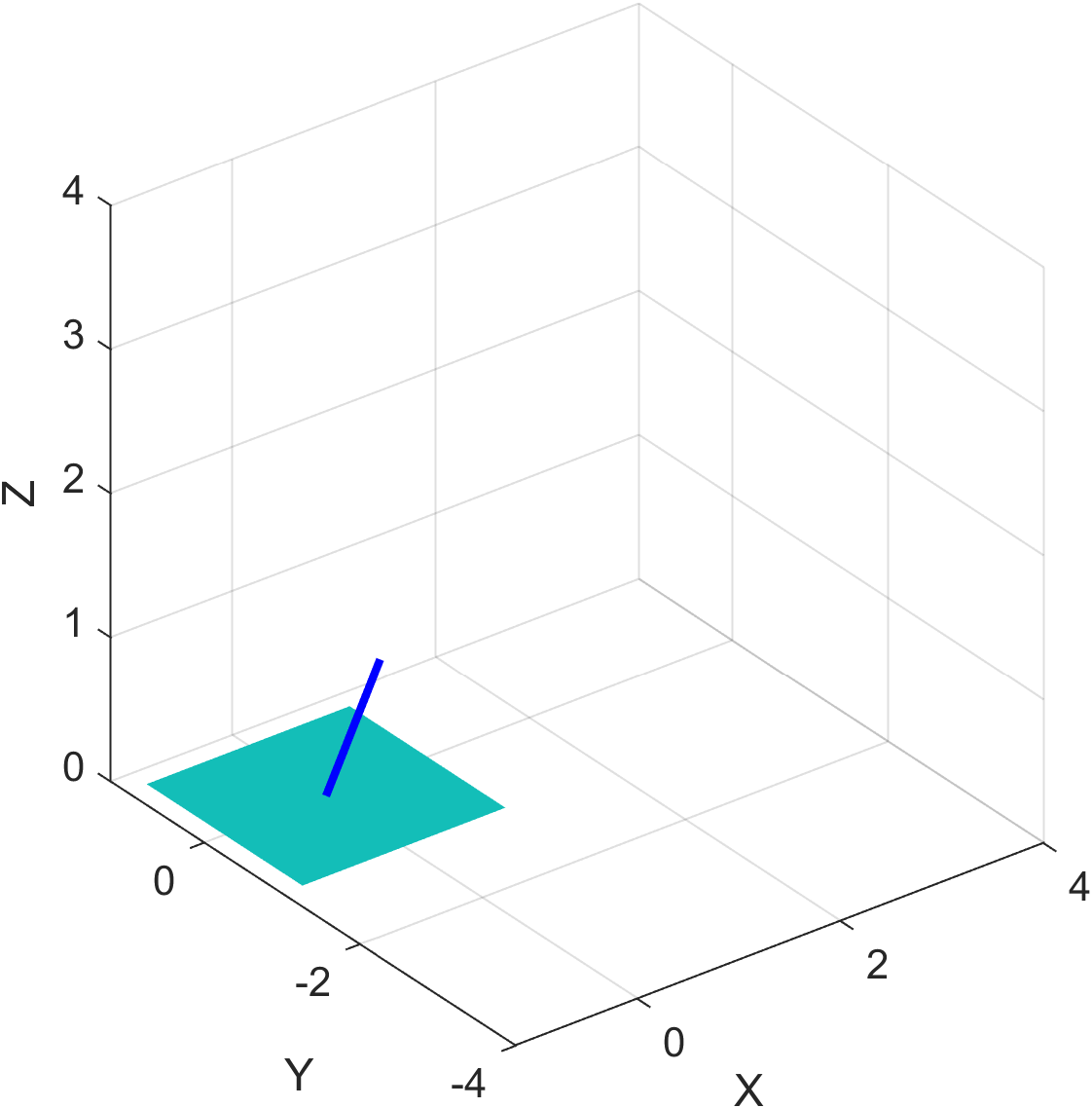}{t = 0 ms} 
        \subgraphics{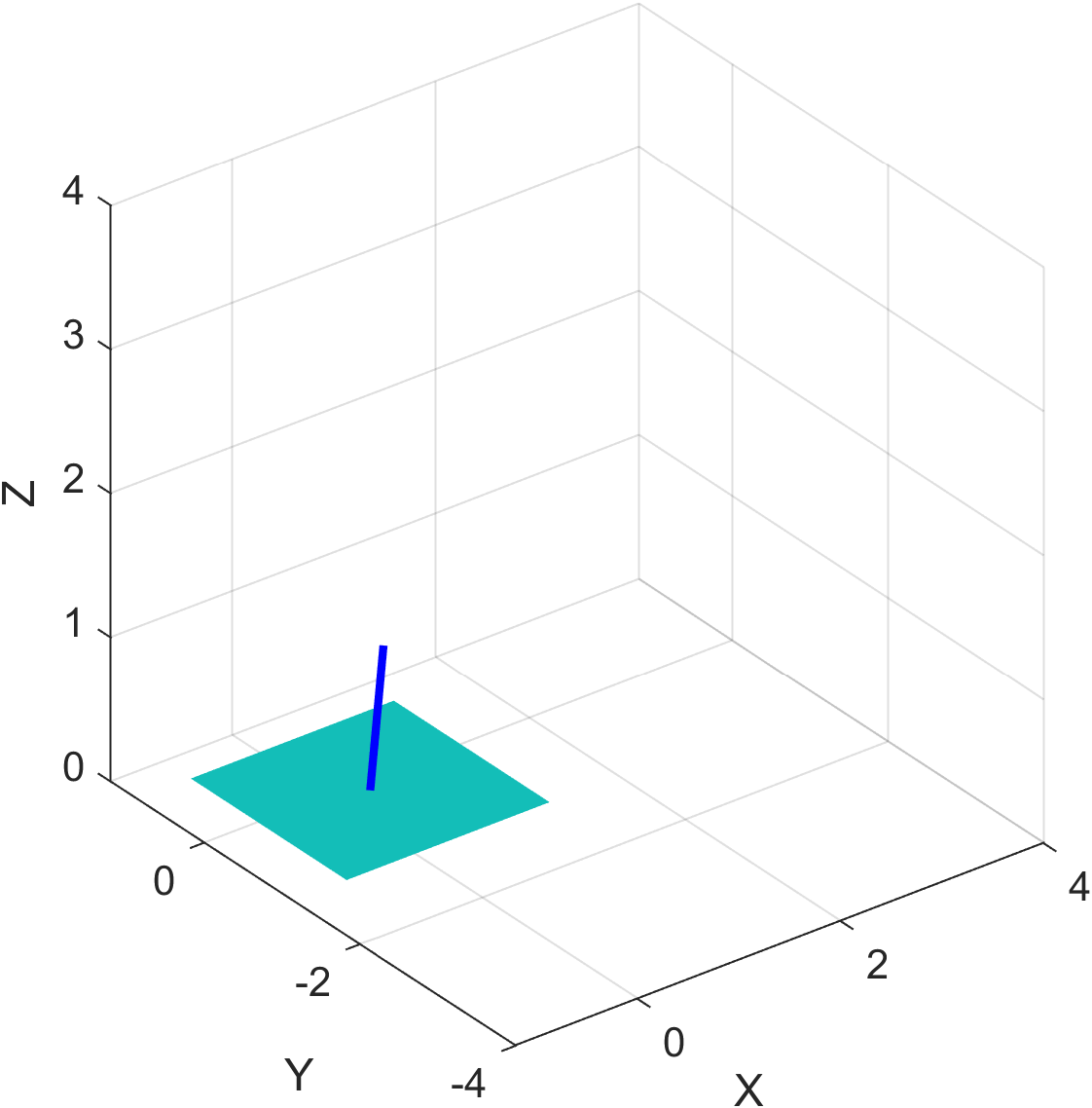}{t = 300 ms} 
        \subgraphics{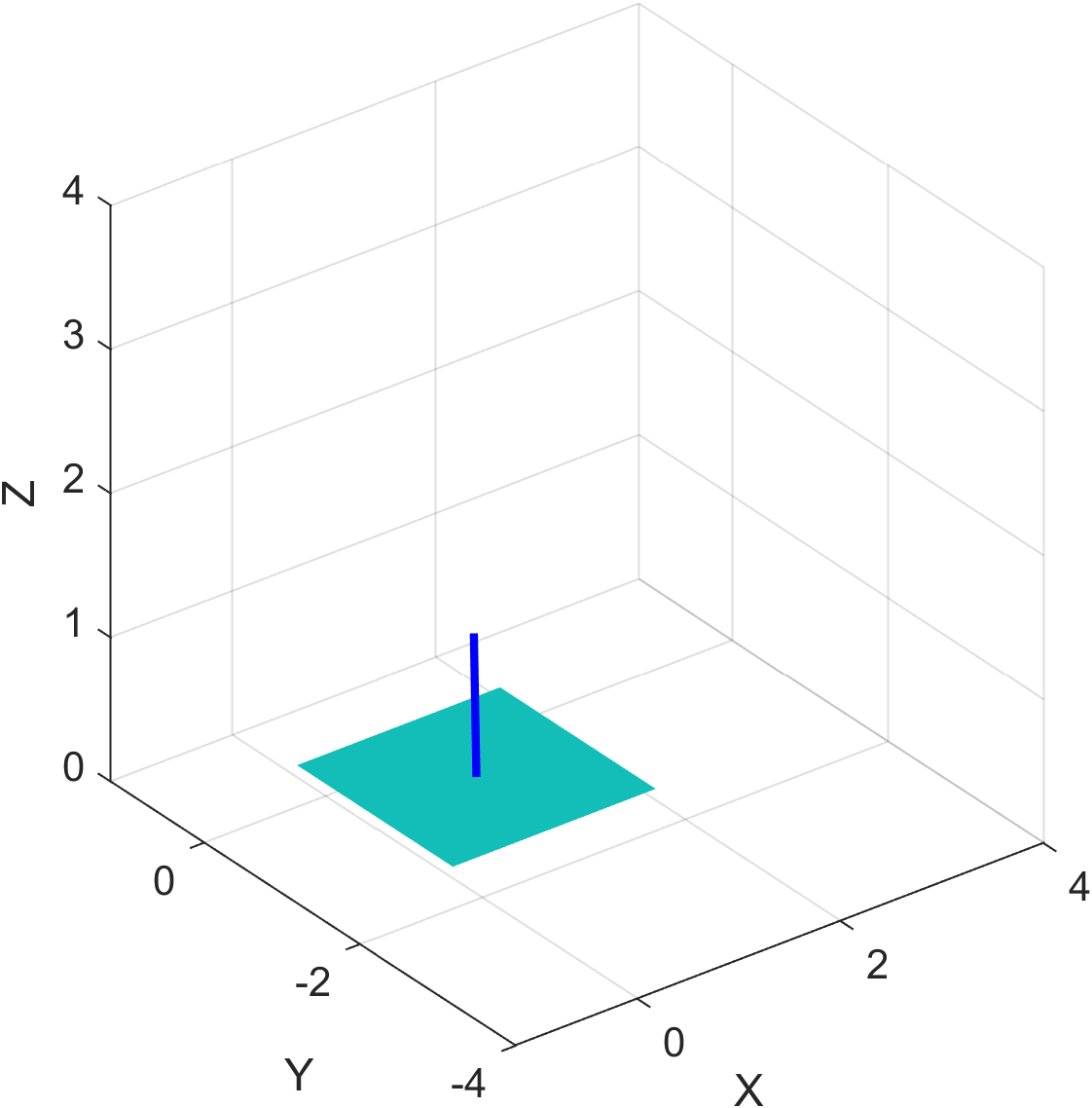}{t = 1100 ms} 
        \caption{Inverted pendulum stabilization : optimal PID control parameters}
        \label{fig:inverted_pendulum_anim_opt}
\end{figure}

\begin{figure}[htbp]
     \centering
     \begin{subfigure}[b]{\textwidth}
         \centering
         \includegraphics[width=\textwidth]{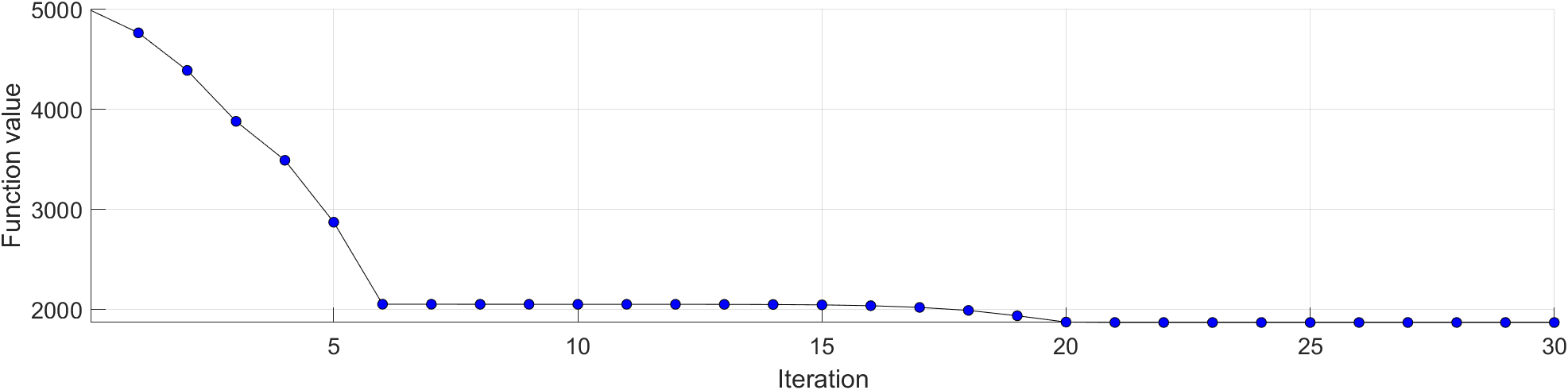}
         \caption{Convergence of objective function for inverted pendulum stabilization}
         \label{fig:inv_pend_obj_func}
     \end{subfigure}
     \hfill
     \begin{subfigure}[b]{\textwidth}
         \centering
         \includegraphics[width=\textwidth]{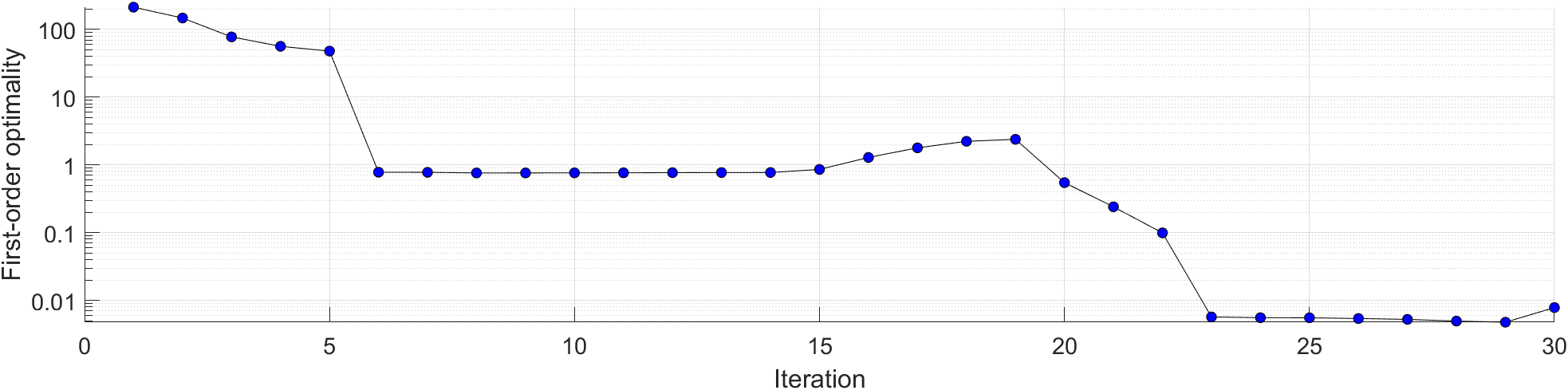}
         \caption{First-order optimality of objective gradient for inverted pendulum stabilization}
         \label{fig:inv_pend_obj_grad}
     \end{subfigure}
     \hfill
    \caption{Inverted pendulum optimization convergence}
    \label{fig:inverted_pendulum_convergence}
\end{figure}

This can be seen graphically in Figure \ref{fig:inverted_pendulum_anim} and Figure \ref{fig:inverted_pendulum_anim_opt}, which show the system state at various simulation times before and after optimization. The difference in the stabilization performance can be clearly observed through the position of the pole against the timestamps. Figure \ref{fig:inverted_pendulum_convergence} shows the convergence of the optimization algorithm. The top subplot is the value of the objective function plotted against the optimization iterations. The optimization process achieved greater than 60\% reduction in the objective function. 

\begin{figure}
    \centering
    \begin{subfigure}{0.49\linewidth}
        \centering
        \includegraphics[width=\linewidth]{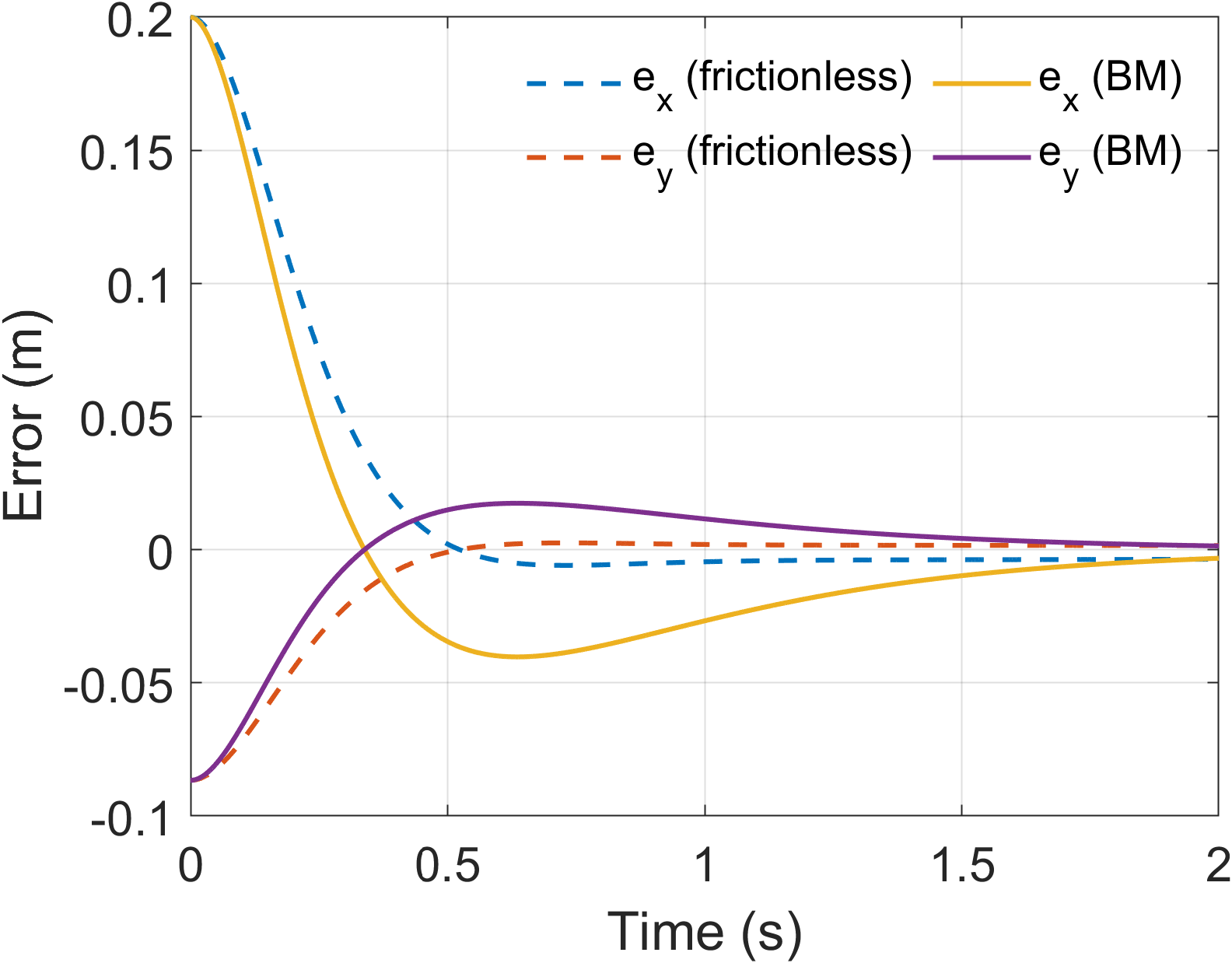}
        \caption{Frictionless Surface}
    \end{subfigure}
    \hfill
    \begin{subfigure}{0.49\linewidth}
        \centering
        \includegraphics[width=\linewidth]{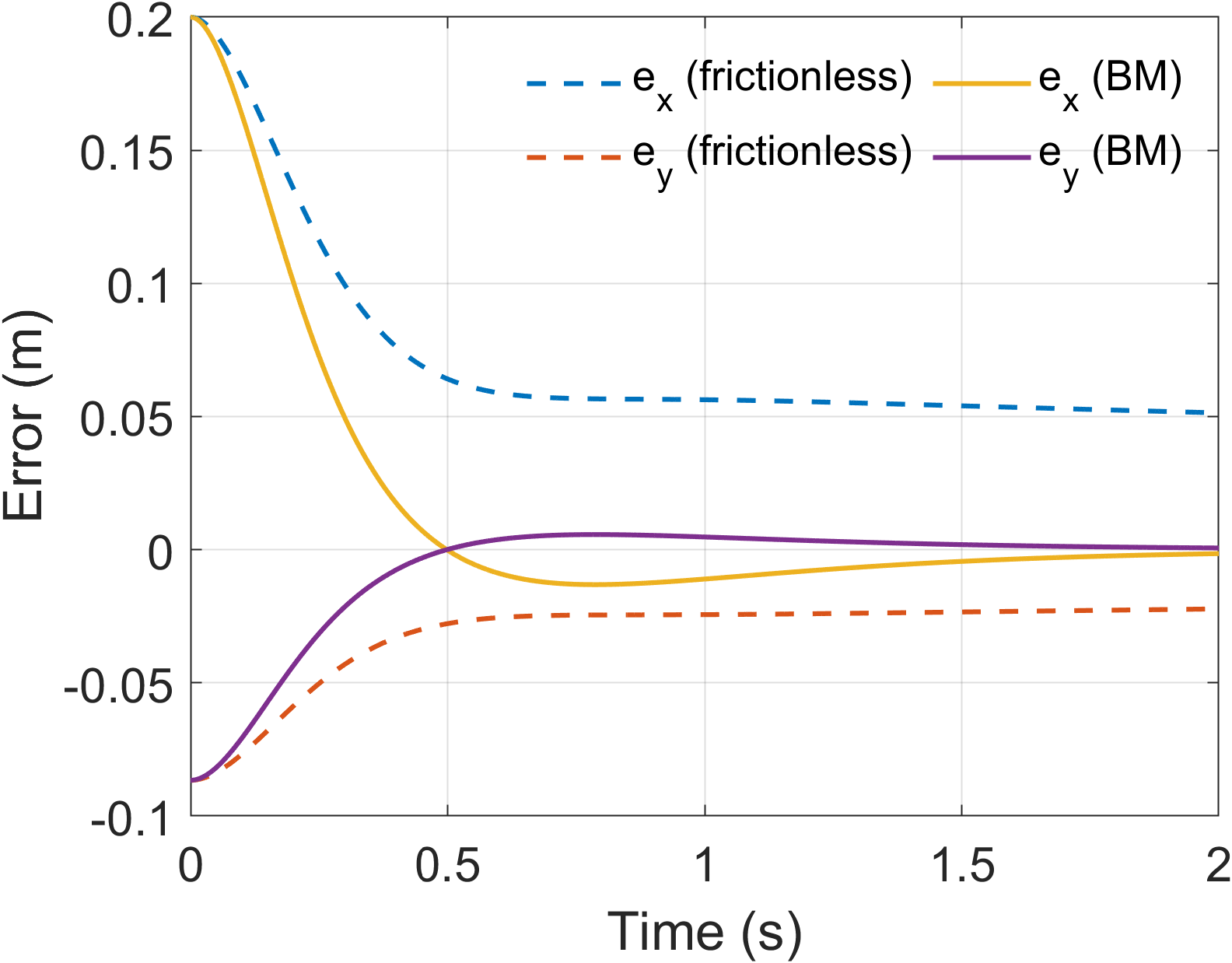}
        \caption{Surface with Gonthier Friction}
    \end{subfigure}
    \caption{Friction versus frictionless optimization error dynamics}
    \label{fig:error_perf_comp}
\end{figure}

Before transitioning to a different case study, it is vital to understand the importance of modeling friction, especially for control problems. Let us suppose that we decide to ignore the friction after making the observation that its magnitude is less than 10\% of the maximum actuator force. The optimization of control parameters is then carried out without any consideration of friction and the control parameters obtained are used in simulation which has friction. Figure \ref{fig:error_perf_comp} compares the stabilization error of the system under such assumptions against previously obtained control parameters where friction was considered during optimization.

\begin{figure}
    \centering
    
    \begin{subfigure}{0.49\textwidth}
        \centering
        \includegraphics[width=\linewidth]{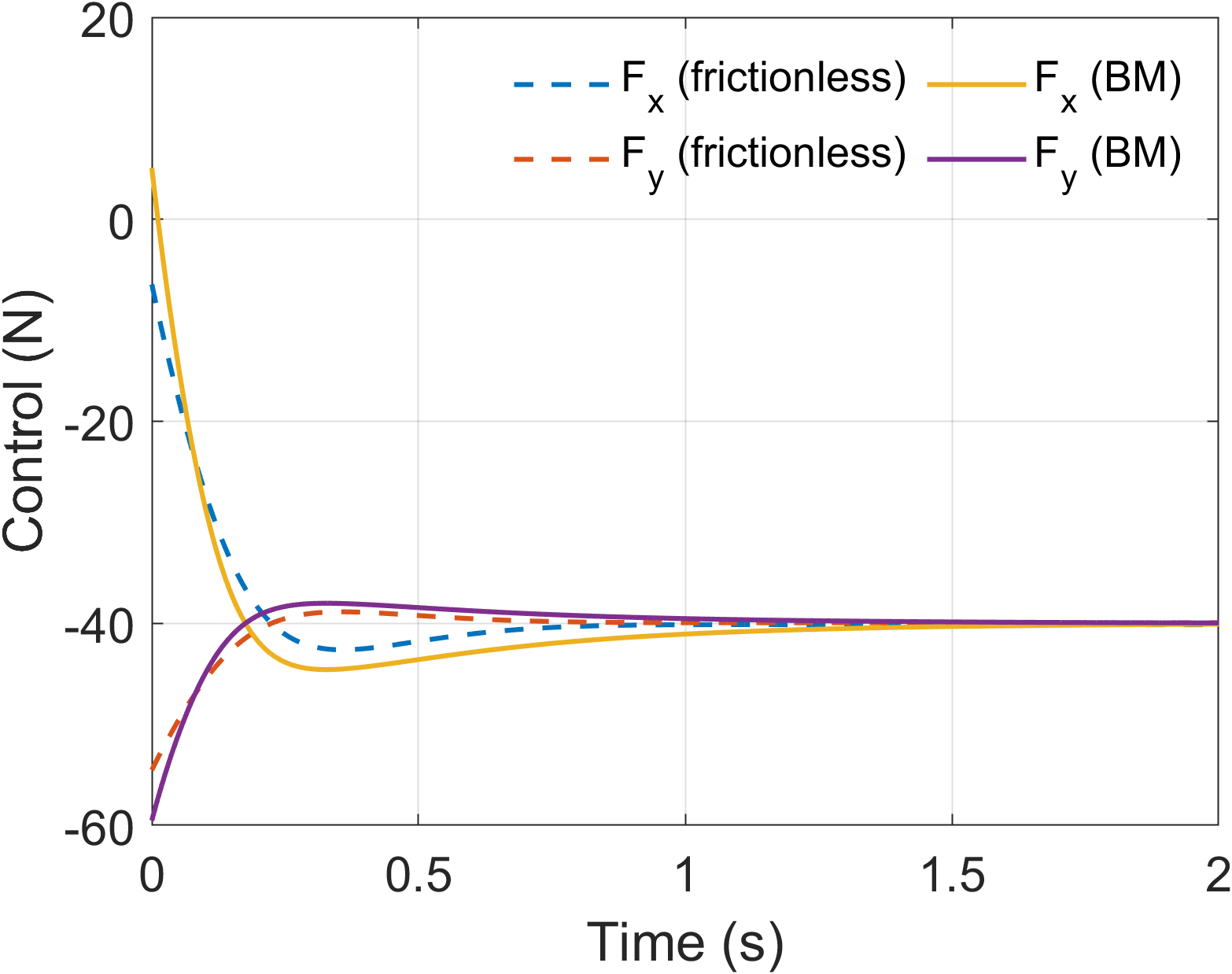}
        \caption{Frictionless Surface}
    \end{subfigure}
    \hfill
    \begin{subfigure}{0.49\textwidth}
        \centering
        \includegraphics[width=\linewidth]{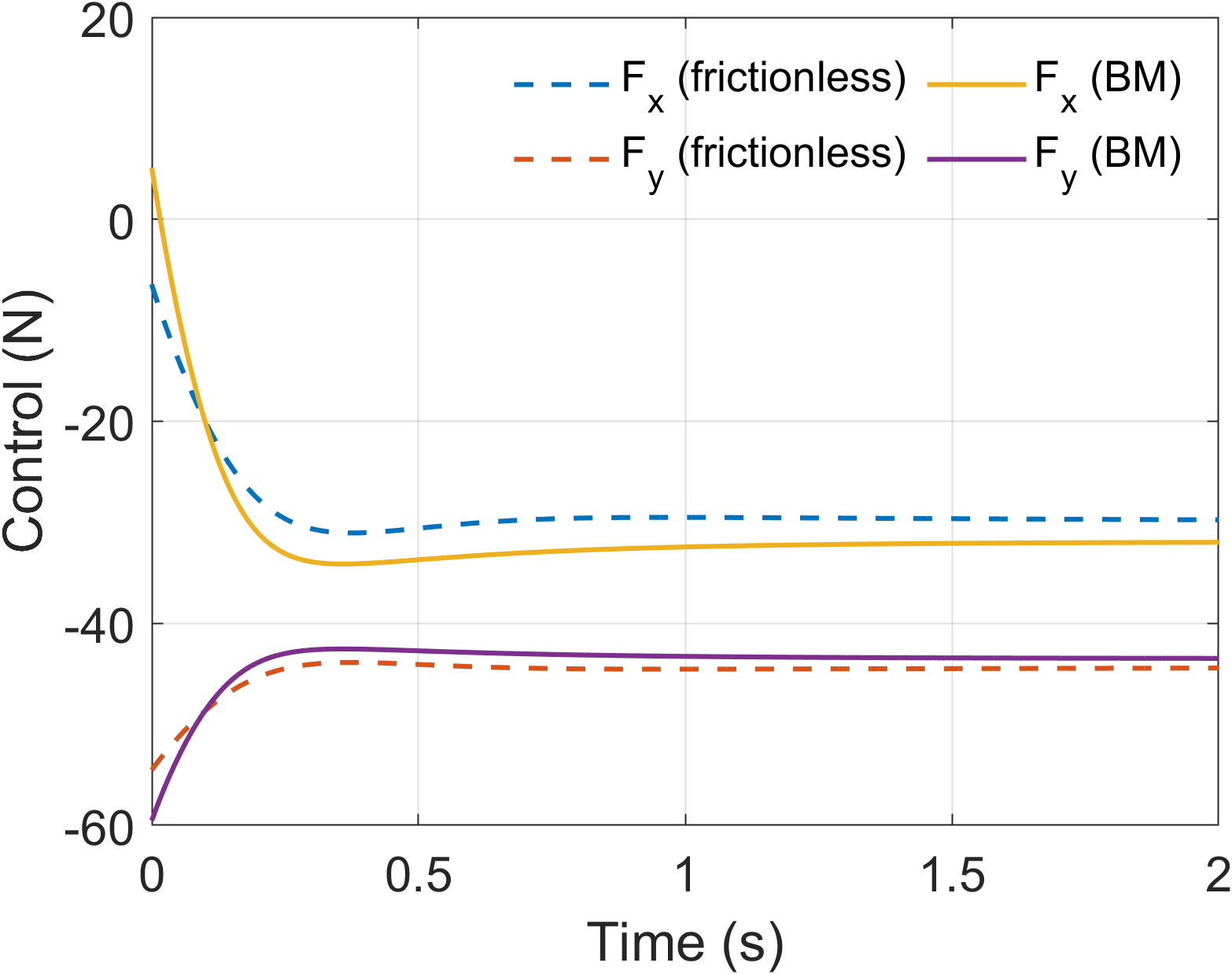}
        \caption{Surface with Gonthier Friction}
    \end{subfigure}
    \caption{Friction versus frictionless optimization control forces}
    \label{fig:control_perf_comp}
\end{figure}

The \textit{dotted} lines represent the optimization model where friction was not considered and the \textit{solid} lines represent the model with friction. If the real world friction was zero, in other words, the surface on which the plate moves is frictionless, the model without friction performs much better in terms of settling time. Figure \ref{fig:control_perf_comp} plots the actuator forces required for stabilization of these models. The model where friction was not considered during optimization achieves stabilization with lower control force magnitudes in comparison to the model where friction was considered during optimization.

Figures \ref{fig:error_perf_comp}(b) and \ref{fig:control_perf_comp}(b) plot the dynamics for a \textit{real-world} setting where the friction has been modeled using the  Gonthier friction model. It also employs different values for static and dynamic friction coefficients in comparison to the Brown McPhee model utilized for optimization ([0.5, 0.4] instead of [0.4, 0.3]). It is evident that the model where friction was not considered during optimization has worse performance than the one where friction was considered. The optimization process for the model without friction converged onto a local optimum [419.42, 30.30, 65.95] instead of the previous optimal  [563.11, 626.57, 90.68]. The optimization ignored the integral gain of the PID controller, rendering it effectively a PD controller. Thereby we see an offset error that does not seem to go away. This occurred because the optimization aimed to minimize control effort without knowledge of the need for integral gain to counteract friction. As a result, it converged to an optimal point with negligible integral gain. The friction force although comparatively small in magnitude to the control forces, has a significant influence on dynamics and thereby cannot be ignored. This study also justifies that the friction model used in the optimization model need not be an exact representation of the friction encountered in real-world.
\subsection{Centrifugal governor mechanism}
\begin{figure}[htbp]
    \centering
    \includegraphics[width=0.75\textwidth]{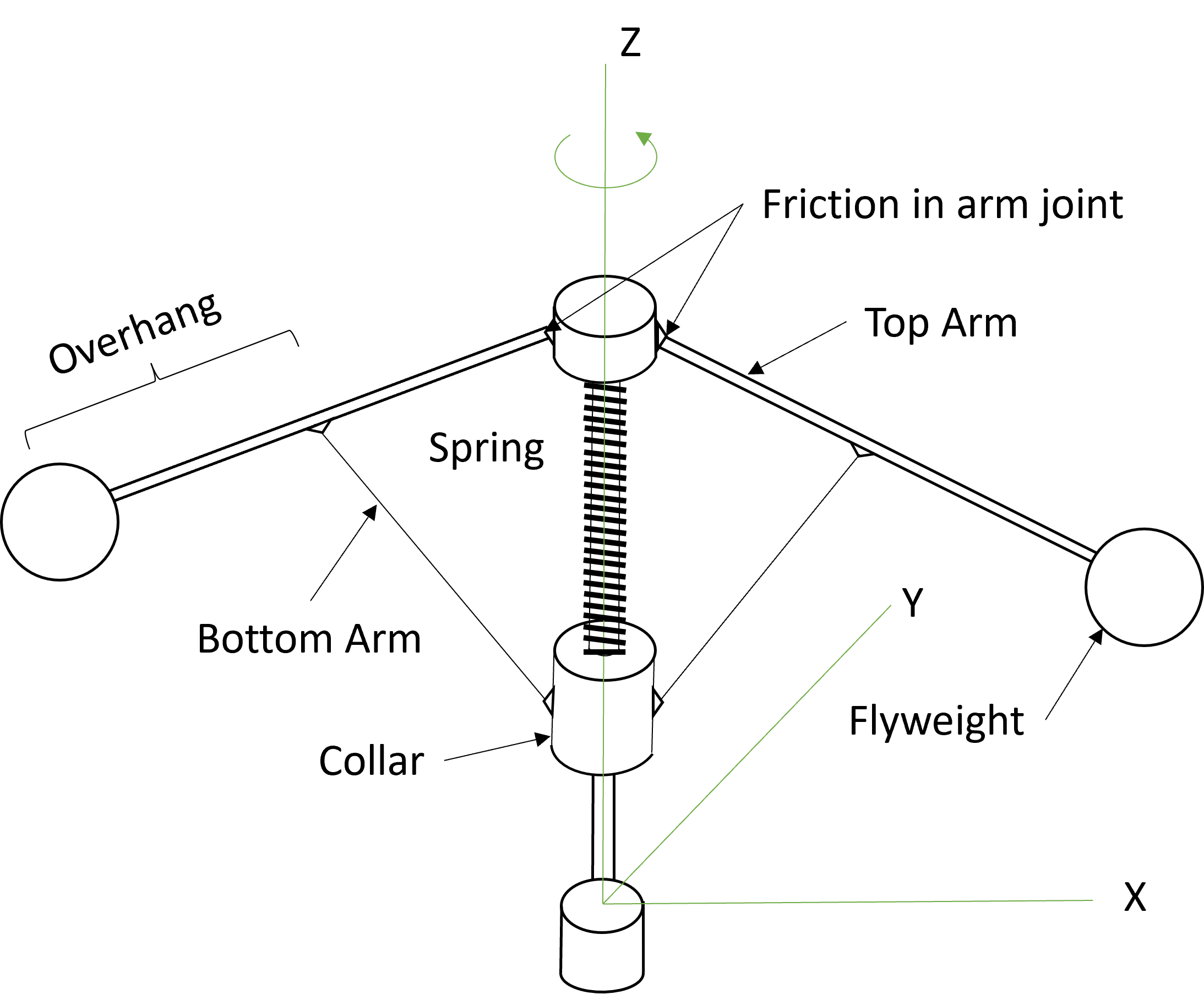}
    \caption{Flyball governor schematic}
    \label{fig:flyball-gov-schematic}
\end{figure}

A centrifugal governor, also known as a flyball governor is typically used to maintain the speed of a combustion engine by regulating the flow of fuel or working fluid. Several commercial applications such as diesel generators and lawn mowers use centrifugal governors. Designing governors for target engine speeds is well understood, making this mechanism an excellent benchmark example for testing the optimization methodology. This mechanism is a type of servo-mechanical proportional controller and its analysis as a dynamic system is not trivial.  Figure \ref{fig:flyball-gov-schematic} shows an example of a type of flyball governor. 

The flyweights attached to the arms are pivoted on a rotating shaft. The collar is attached to the top arm via links and is constrained to translate along the shaft. The entire mechanism rotates with the same angular velocity, say $\omega$, which is proportional to the engine speed. As $\omega$ decreases due to an increase in engine load, the centrifugal forces on the weights decreases causing them to move inward thereby decreasing the collar's height and vice versa. The collar can be connected the throttle valve of the engine to control the fuel flow and thereby the engine speed. This provides the said governing action for the engine. A key component of such governors is the translational spring-damper (TSD) which opposes the motion of the collar to provide a damped dynamic response. For the purpose of this study the TSD stiffness $k_s = 1000$ N/m and damping of $c_s = 50.0$  N-s/m has been used. The servo-mechanism torque or torque at the pillar $\tau_{\text{pillar}}$ has been modelled using a linear proportional controller governed by the following equation:
\begin{align}
    \tau_{\text{pillar}} = K_p (h_0 - z_2),
\end{align}
where $K_p = 200.0$ N-m/m is the proportional gain in torque, $h_0= 0.1 \, \text{m}$ is some predefined height of the collar where the torque would be zero and $z_2$ is the dynamic height of the collar obtained in simulation. The system is highly nonlinear, especially if friction effects are considered. Therefore it is a good benchmark problem to test the methodology. Such mechanisms are severely neglected in terms of maintenance. Hence, it is imperative to ensure that the mechanism functions as desired even if improper maintenance, such as lack of adequate lubrication, leads to friction in this mechanism. 

\begin{table}[htbp]
    \centering
    \caption{Flyball governor: number of equations}
    \begin{tabular}{@{}lc@{}}
    \toprule
    \textbf{Component} & \textbf{Value} \\
    \midrule
    Number of bodies & $6$ \\
    States per body & $7$ \\
    Total differentiable variables for dynamics & $6\times 7 = 42$ \\
    First-order equations of motion & $2 \times 42 = 84$ \\
    Degrees of freedom & $2$  \\
    Lagrange multipliers/Constraints & $42-2 = 40$ \\
    Total dynamic equations & $84+40 = 124$ \\
    Number of free-variables (parameters) & $7$\\
    Total number of sensitivities & $7\times 124 = 868$\\
    Total differential-algebraic equations & $124+868=992$\\
    Total objective function(s) & $1$\\
    Total objective function gradients & $7$\\
    \bottomrule
    \end{tabular}
    \label{tab:governor_DAE}
\end{table}

In terms of modeling, the system contains 6 bodies (2 top links, 2 bottom links, a vertical pillar which rotates about the Z axis, and a sliding collar). This mechanism is axisymmetric and can also be modelled with 4 bodies by adding the symmetrical centrifugal forces in the generalized force vector. The friction in this system has been modeled in the revolute joint of the top arm which holds the flyweights. This joint is bound to experience high constraint and inertial forces due to the rotation of the governor. Hence, due to improper maintenance, this joint will experience increased frictional torque. The friction at the collar-pillar translational joint will be low, since the centrifugal forces cancel out, leading to negligible constraint reaction forces at that joint. The objective of this particular example is to modify the governor design to achieve a certain desired stable speed, in the presence of friction.
\begin{subequations}
\begin{align}
    \min_{\bm\rho}\psi &= \int_{t_0}^{t_f} (\omega_z - 15.0)^2 \text{d}t,\\
    \text{where}\quad\bm\omega &=  \left[ \begin{array}{ccc}
          \omega_x, & \omega_y, & \omega_z
    \end{array}\right]^{\text{T}}
\end{align}
\label{eq:gov_obj}
\end{subequations}
The angular velocity for any body $i$ is purely a function of its Euler parameters $\mathbf{p}_i$. We have $\bm\omega_i = \mathbf{E}(\mathbf{p}_i) \dot{\mathbf{p}_i}$. The objective function can be expressed mathematically as shown in Equation (\ref{eq:gov_obj})a.

\begin{figure}[htbp]
    \centering
    \includegraphics[width=\textwidth]{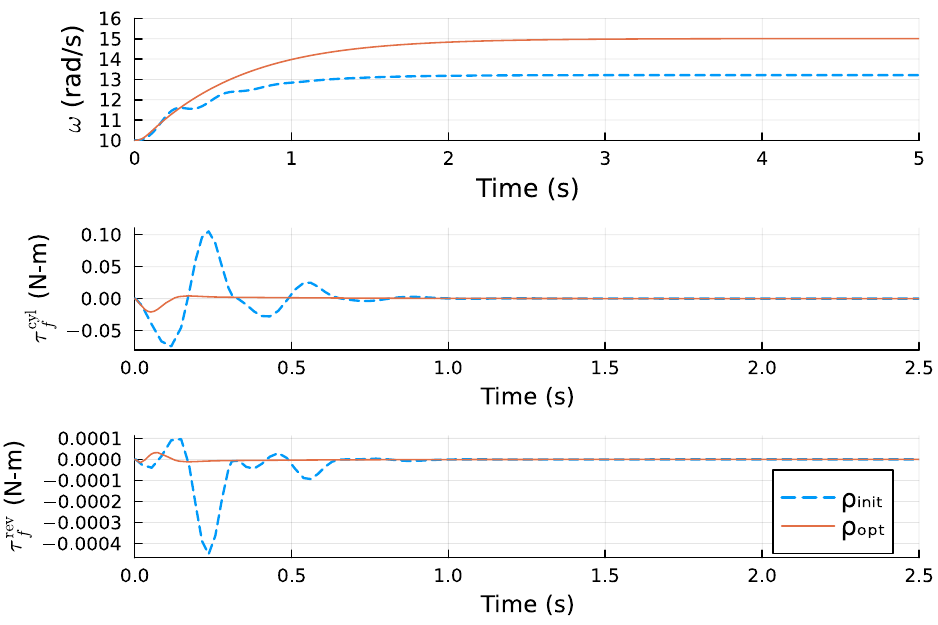}
    \caption{Flyball governor dynamic plots}
    \label{fig:flyball_gov_dyn}
\end{figure}

The design parameters for this case study are shown in Table \ref{tab:governor_parameters}. As it can be seen in Figure \ref{fig:flyball-gov-schematic}, the bottom arm of the governor does not connect with the top arm at the very end resulting in an overhang. This gives the flyweights leverage to lift the collar against the spring force. The ratio of the overhang to the total length of the top arm is another design parameter for the system. Parameters like pillar height and collar outer radius may not have a substantial effect on the optimization.

\begin{table}[htbp]
    \centering
    \caption{Flyball governor parameter values}
    \begin{tabular}{@{}lllll@{}}
        \toprule
        \textbf{Parameter} & \textbf{Description} & \textbf{Initial Est.} & \textbf{Bounds} & \textbf{Optima}\\
        \midrule
        $\rho_1$ & Top arm length & 0.135 m & [0.1 m, 0.15 m] & 0.1247 m\\
        $\rho_2$ & Bottom arm length & 0.08 m & [0.06 m, 0.1 m] & 0.0853 m\\
        $\rho_3$ & Overhang ratio & 0.3 & [0.2, 0.4] & 0.20\\
        $\rho_4$ & Top arm to Pillar offset & 0.025 m & [0.02 m, 0.03 m] & 0.02 m\\
        $\rho_5$ & Collar outer radius & 0.025 m & [0.02 m, 0.03 m] & 0.02 m\\
        $\rho_6$ & Pillar height & 0.225 m & [0.2 m, 0.25 m] & 0.2 m\\
        $\rho_7$ & Flyweight mass & 0.03 kg & [0.02, 0.04] & 0.0242 kg\\
        \bottomrule
    \end{tabular}
    \label{tab:governor_parameters}
\end{table}

Figure \ref{fig:flyball_gov_dyn} shows the initial and optimized trajectory of the governor speed. The optimization achieves the target speed of 15 rad/s. Almost all parameters have decreased in magnitude, except for the bottom arm length. The correlation of the top and bottom arm length to pendulum speed is peculiar. Moving the centrifugal masses outward should decrease the target speed since the same centrifugal force can be achieved at a lower speed. However, changing the top arm length and overhang ratio changes the geometry of the mechanism, specifically the joint location for the top and bottom arms. Hence, to maintain this constraint, the length of the bottom arm increased by a small amount even though the top arm length decreased in the optimal design.

\begin{figure}[htbp]
    \centering
    \includegraphics[width=0.75\textwidth]{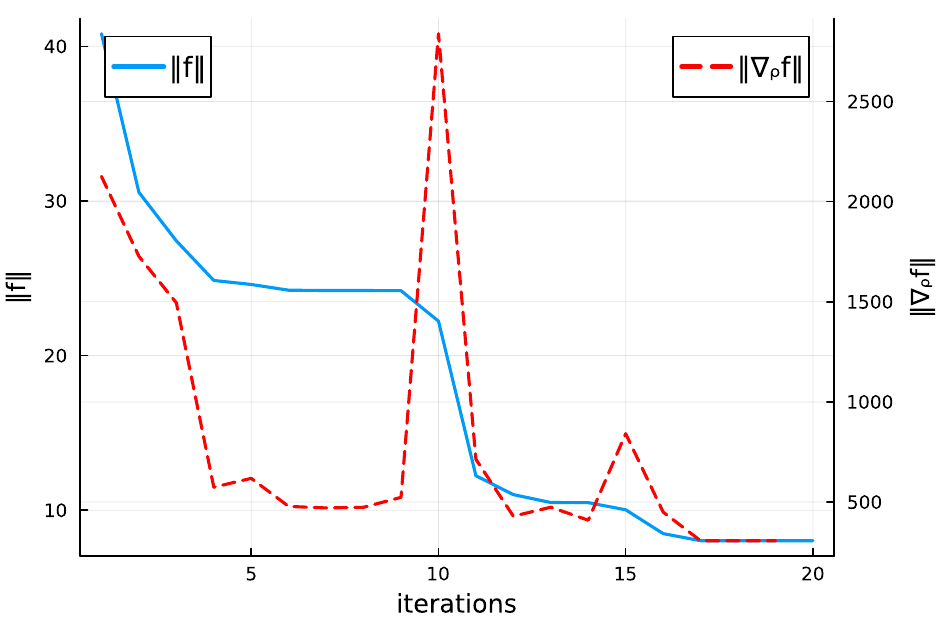}
    \caption{Flyball governor convergence}
    \label{fig:flyball_gov_convergence}
\end{figure}

Figure \ref{fig:flyball_gov_convergence} shows the optimization converging to a local minimum. It is important to mention that the performance of this particular optimization case study was far from ideal due to the high number of differential equations and sub-optimal memory management. This study is intended as a proof of concept of the optimization methodology and improvements to the execution speed and computational efficiency are planned in the near future.
\subsection{Spatial slider-crank mechanism}
\label{sec:test-slider-crank}
\begin{figure}[htbp]
    \centering
    \includegraphics[width=\textwidth]{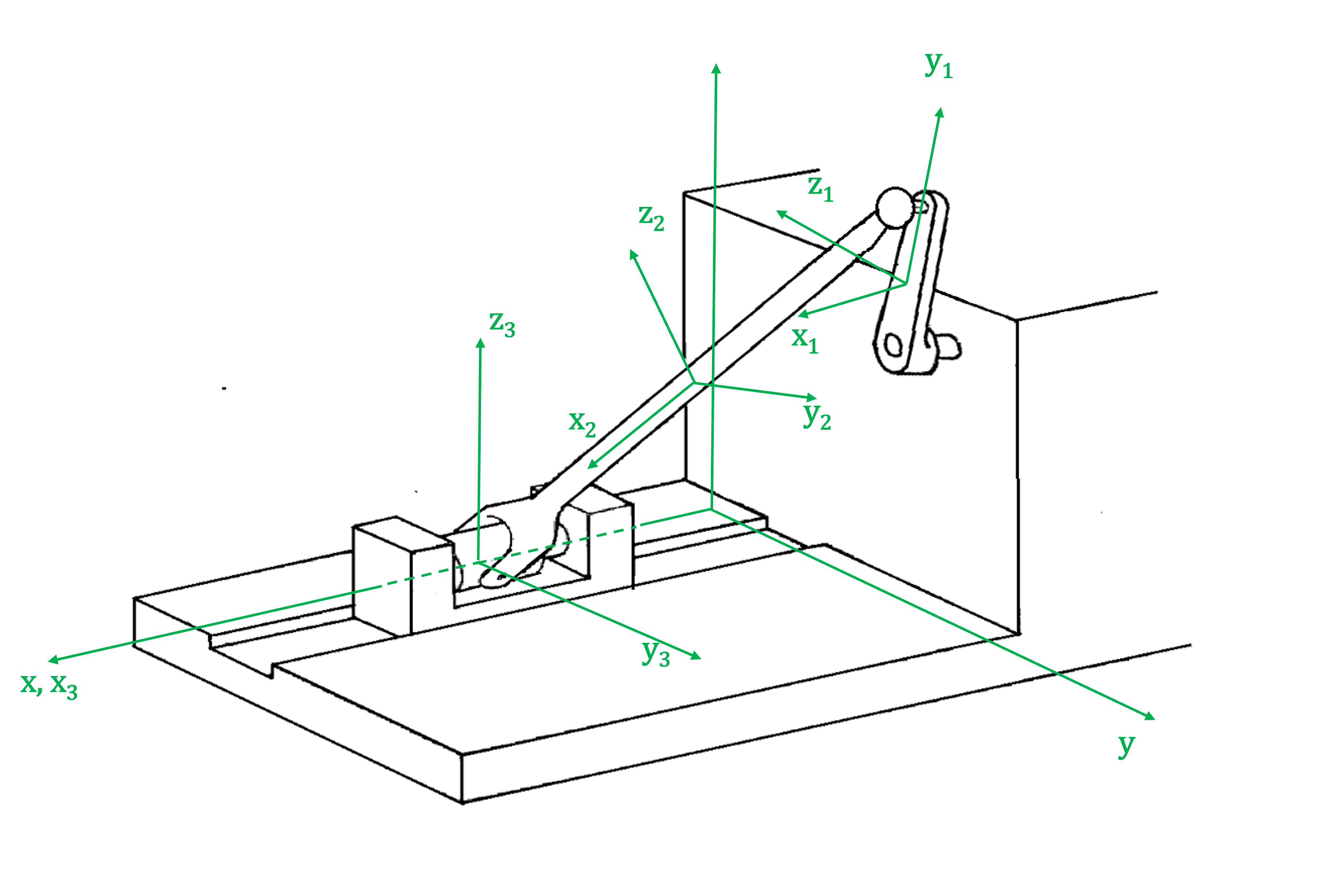}
    \caption{Slider-crank mechanism schematic}
    \label{fig:slider-crank_schematic}
\end{figure}

The slider-crank mechanism is a multibody system which converts rotational motion at the crank into oscillating translational motion at the slider. This mechanism is ubiquitous and is used in combustion engines, and various manufacturing processes, such as shaper machines, sheet metal punching machines and shearing machines. For the purpose of this study, a spatial version of the mechanism is used, as shown in Figure \ref{fig:slider-crank_schematic}. The model is derived from Haug (1989) \cite{Haug1989}, which contains the kinematic analysis of the mechanism.

In a previous paper \cite{Verulkar2022}, the sensitivity analysis of this mechanism was presented and the dynamic response was benchmarked using two friction models. It was also demonstrated that the sensitivities of multibody systems with Brown and McPhee friction, behave like hybrid-dynamic systems, by displaying abrupt jumps. This paper will build upon the aforementioned study and illustrate how the optimization methodology presented herein can be applied for co-design optimization. Table \ref{tab:slider-crank_noe} highlights the number of equations to be solved for every optimization iteration.

\begin{table}[htbp]
    \centering
    \caption{Slider-crank : number of equations}
    \begin{tabular}{@{}lc@{}}
    \toprule
    \textbf{Component} & \textbf{Value} \\
    \midrule
    Number of bodies & $3$ \\
    States per body & $7$ \\
    Total differentiable variables for dynamics & $3\times 7 = 21$ \\
    First-order equations of motion & $2 \times 21 = 42$ \\
    Degrees of freedom & $1$  \\
    Lagrange multipliers/Constraints & $21-1 = 20$ \\
    Total dynamic equations & $42+20 = 62$ \\
    Number of free-variables (parameters) & $5$\\
    Total number of sensitivities & $5\times 62 = 310$\\
    Total differential-algebraic equations & $62+310=372$\\
    Total objective function(s) & $1$\\
    Total objective function gradients & $5$\\
    \bottomrule
    \end{tabular}
    \label{tab:slider-crank_noe}
\end{table}

The objective of this study is to design the system and control parameters such that the crank spins at a given constant rotational velocity through a proportional control mechanism. Thus, the control torque on the crank $\tau_{\text{crank}}$ can be represented by the following equation:

\begin{align}
    \tau_{\text{crank}} = (\omega - \omega_0)K_p,
\end{align}

where $\omega$ is the current dynamic angular velocity of the crank $\omega_0 = -10.0$ rad/s is some predefined speed for the crank rotation and $K_p$ is the tunable proportional controller for crank torsional actuator. The objective function used in the study is:

\begin{align}
    \psi = \int_{t_0}^{t_f} \left[(\omega-\omega_0)^2 + 0.01 \tau_{\text{crank}}^2\right]\,\text{d}t = \int_{t_0}^{t_f} \left[(\omega-\omega_0)^2 (1 + 0.01 K_p^2)\right]\,\text{d}t.
    \label{eq:slider-crank_obj_fcn}
\end{align}

\begin{table}[htbp]
    \centering
    \caption{Slider-crank parameter values}
    \begin{tabular}{@{}lllll@{}}
        \toprule
        \textbf{Parameter} & \textbf{Description} & \textbf{Initial} & \textbf{Bounds} & \textbf{Optima}\\
        \midrule
        $\rho_1$ & Crank length & 0.08 m & [0.06 m, 0.1 m] & 0.06 m\\
        $\rho_2$ & Connecting rod length & 0.3 m & [0.2 m, 0.4 m] & 0.2 m\\
        $\rho_3$ & Slider length & 0.05 & [0.04 m, 0.06 m] & 0.0467 m\\
        $\rho_4$ & Slider width & 0.025 m & [0.020 m, 0.03 m] & 0.02 m\\
        $\rho_5$ & Proportional gain & 1.0 & [-5.0, 5.0 ] & 1.0495 \\
        \bottomrule
    \end{tabular}
    \label{tab:slider-crank_parameters.}
\end{table}

\begin{figure}[htbp]
    \centering
    \includegraphics[width=\textwidth]{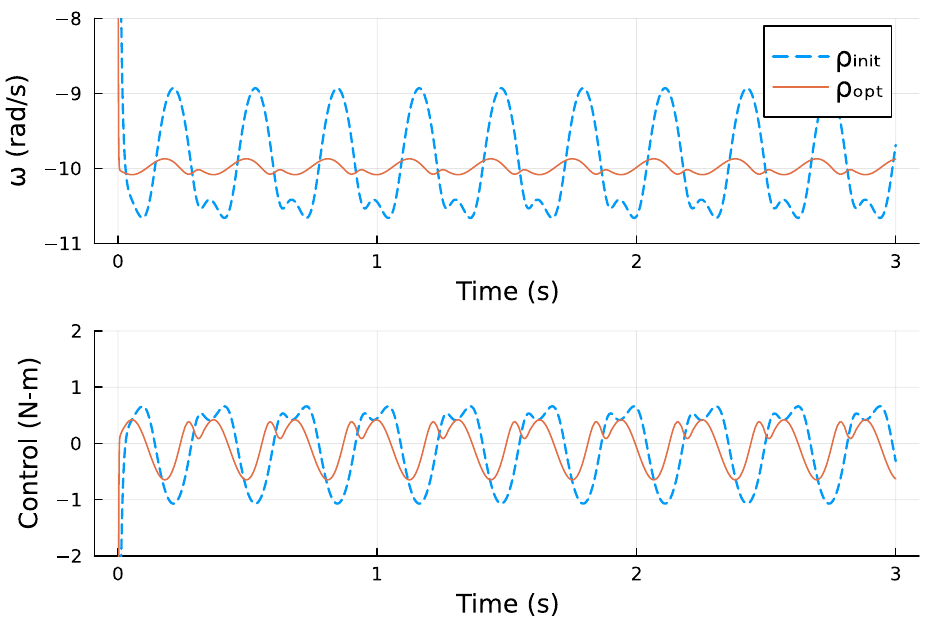}
    \caption{Slider-crank dynamics}
    \label{fig:slider-crank_dynamics}
\end{figure}

Figure \ref{fig:slider-crank_dynamics} shows the dynamic plots before and after optimization. It can be seen that the crank is continuously rotating at approximately 10 rad/s with the optimized parameters. The maximum magnitude of control required is also lower. This is a typical advantage of co-design optimization. It will often be the case that the control effort can be substantially reduced if appropriate design is chosen as seen in this example. 

\begin{figure}[htbp]
    \centering
    \includegraphics[width=\textwidth]{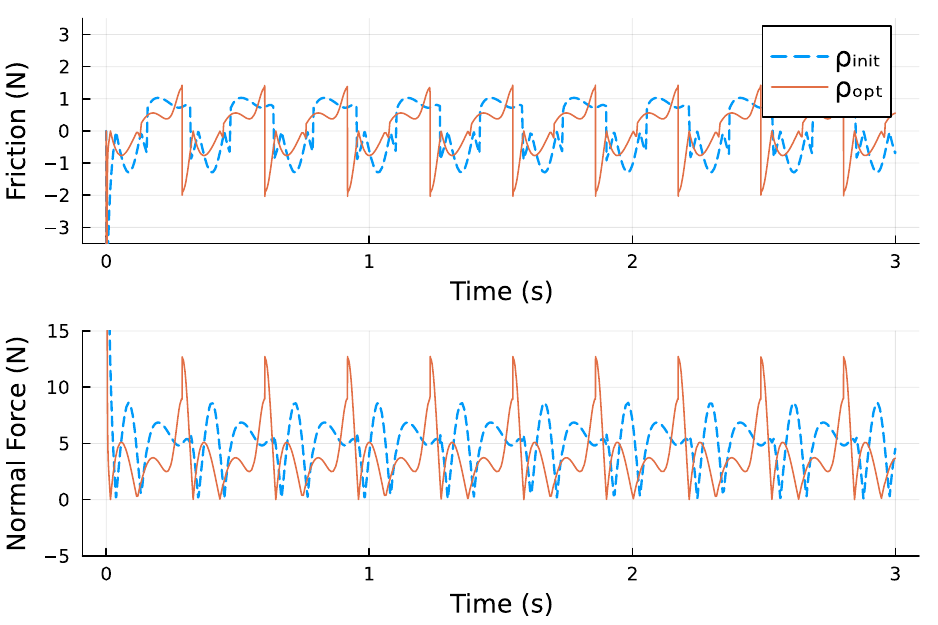}
    \caption{Slider-crank normal and friction forces}
    \label{fig:slider-crank_friction}
\end{figure}

Figure \ref{fig:slider-crank_friction} shows the friction and effective normal reaction at the translational joint. As it can be observed, the initial response of the system was highly oscillatory which was due to high jerks when the crank was at the bottom most position. 

\begin{figure}[htbp]
    \centering
    \includegraphics[width=\textwidth]{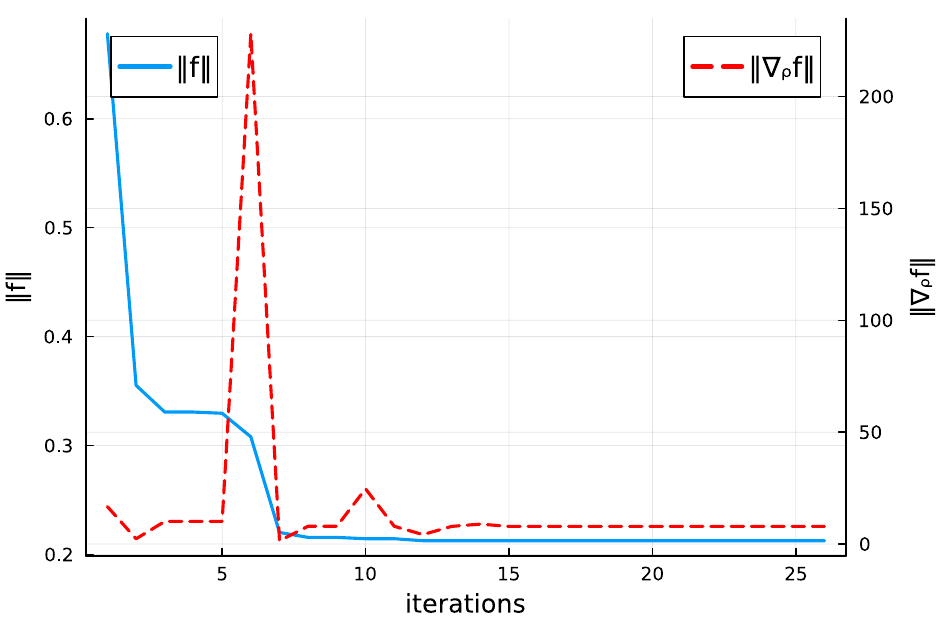}
    \caption{Slider-crank convergence}
    \label{fig:slider-crank_convergence}
\end{figure}

Figure \ref{fig:slider-crank_convergence} shows the convergence of the optimization problem. The objective function has decreased by more than three times in comparison to the initial estimate of design and control parameters.

\section{Conclusions}
\label{sec:conclusions}
This paper covered the direct sensitivity methodology for optimization of multibody systems with friction. It was demonstrated how systems with friction can be practically modeled using singular mass-matrix form of differential-algebraic equations that are compatible with standard open-source solvers. Sensitivity analysis through direct differentiation of such dynamic equations can be solved through explicit solvers with adaptive stiffness control. The examples examined in this study underscore the versatility of this methodology in both control and design parameter optimization while demonstrating the advantages of co-design. The methodology is also applicable to control shaping and dynamic estimation problems, however adjoint sensitivity approach would be preferable in this case due to the large number of parameters. It also gives a primer on a revised version of MBSVT that will be made open-source after desired performance improvements are made. At the moment, since many of the optimizations have not been implemented, the MBSVT-Julia package is not competitive against other commercial tools such as CasADi/IPOPT in terms of dynamic system optimization time for a large number of states and parameters. As a future scope of this work, more efficient sparse-matrix implementations need to be executed with non-allocating functions to further speed up the execution time. Additionally, general adjoint sensitivity approaches for optimal control based on calculus of variations plan to be included.

\nomenclature[01]{$nb$}{Number of bodies in the system.}
\nomenclature[02]{$n$}{Number of generalized coordinates. $7 nb$ for reference point coordinates with Euler parameters.}
\nomenclature[03]{$\mathbf{q}\in \mathbb{R}^{n}$}{Vector of generalized coordinates.}
\nomenclature[04]{$\bm{\lambda}\in \mathbb{R}^{m}$}{Vector of Lagrange multipliers.}
\nomenclature[05]{$\bm\rho\in\mathbb{R}^p$}{Vector of system parameters $[\rho_1,\, ...\,,\,\rho_p]^{\text{T}}$.}
\nomenclature[06]{$\mathbf{y}_{\mathbf{x}}$}{Jacobian formed by partial derivatives $\frac{\partial \mathbf{y}}{\partial \mathbf{x}}$.}
\nomenclature[07]{$\dot{x}, \ddot{x}$}{Time derivatives $\frac{\text{d}x}{\text{d}t}$ and $\frac{\text{d}^2 x}{\text{d}t^2}$ respectively.}
\nomenclature[08]{$\bm \Phi(\mathbf{q}, \bm\rho) \in \mathbb{R}^m$}{Vector of $m$ holonomic constraints.}
\nomenclature[09]{$\bm \Phi_{\mathbf{q}}\in \mathbb{R}^{m\times n}$}{The constraint vector Jacobian. Must be of rank $m$.}
\nomenclature[10]{$\mathbf{M}(\mathbf{q}, \bm\rho)\in \mathbb{R}^{n\times n}$}{Generalized mass matrix for multibody systems.}
\nomenclature[11]{$\mathbf{Q}(\mathbf{q}, \dot{\mathbf{q}}, \bm\rho)\in \mathbb{R}^{n}$}{Vector of external generalized forces and torques.}
\nomenclature[12]{$\mathbf{Q}^{Af}(\mathbf{q}, \dot{\mathbf{q}}, \bm\lambda, \bm\rho)\in \mathbb{R}^{n}$}{Vector of generalized frictional forces.}
\nomenclature[13]{$v$}{Magnitude of relative sliding velocity.}
\nomenclature[14]{$v_t$}{Magnitude of transition velocity.}
\nomenclature[15]{$\bm\mu$}{$\left[\mu_d, \quad \mu_s\right]^\text{T}$, wherein $\mu_d$ and $\mu_s$ represent coefficients for dynamic and static friction, respectively.}
\nomenclature[16]{$F_n(\mathbf{q}, \bm\lambda,\bm\rho)$}{Magnitude of normal force of contact.}
\nomenclature[17]{$F_f$}{Magnitude of friction force.}
\nomenclature[18]{$\mathbf{r}_i\in\mathbb{R}^3$}{Position vector in global frame for the origin of $i$\textsuperscript{th} body-fixed reference frame.}
\nomenclature[19]{$\mathbf{u}_x$, $\mathbf{u}_y$, $\mathbf{u}_z$ $\in\mathbb{R}^3$}{Unit axis representing a coodinate system.}
\nomenclature[20]{$(\cdot)$, $(\cdot)^\prime$, $(\cdot)^{\prime\prime}$}{Physical entities represented in the ground, body-fixed, and joint reference frames respectively.}
\nomenclature[21]{$\Tilde{\mathbf{x}}\in\mathbb{R}^{3\times 3}$}{The matrix $[0,\, -x_3,\, x_2;\, x_3,\, 0,\, -x_1;\, -x_2,\, x_1,\, 0]$ formed using the elements of a 3-vector $\mathbf{x}=[x_1;\, x_2;\, x_3]$. For any other 3-vector $\mathbf{y}$, we have $\Tilde{\mathbf{x}}\mathbf{y} = \mathbf{x}\times \mathbf{y}$.}
\nomenclature[22]{$\mathbf{p}_i\in\mathbb{R}^4$}{The Euler parameter vector for the $i$\textsuperscript{th} body-fixed reference frame is expressed as $[e_0\quad\mathbf{e}^{\text{T}}]^{\text{T}}$, where $e_0 = \cos({\chi/2})$, and $\mathbf{e} =\mathbf{u}\sin({\chi/2})$ for a given axis-angle rotation $\mathbf{u}$ and $\chi$.}
\nomenclature[23]{$\mathbf{s}^{\prime}_i\in\mathbb{R}^3$}{Position vector for joint location in the local $i$\textsuperscript{th} body-fixed reference frame.}
\nomenclature[24]{$\mathbf{A}_i\in\mathbb{R}^{3\times3}$}{The rotation matrix in terms of Euler parameters, which denotes the transformation from the $i$\textsuperscript{th} body-fixed coordinate system to the global coordinate system, is given by $\mathbf{A}(\mathbf{p})=(e_0^2-\mathbf{e}^{\text{T}}\mathbf{e})\mathbf{I}+2\mathbf{e}\mathbf{e}^\text{T}+2e_0\Tilde{\mathbf{e}}$.}
\nomenclature[25]{$\mathbf{B}_i\in\mathbb{R}^{3\times4}$}{$2\left[\left(e_0\mathbf{I}+\Tilde{\mathbf{e}}\right)\mathbf{s}^{\prime}_i\quad\mathbf{e}\mathbf{s}^{\prime\text{T}}_i-\left(e_0\mathbf{I}+\Tilde{\mathbf{e}}\right)\Tilde{\mathbf{s}^{\prime}_i}\right]$.}
\nomenclature[26]{$\mathbf{E}_i\in\mathbb{R}^{3\times4}$}{$\left[-\mathbf{e}\quad \Tilde{\mathbf{e}}+e_0\mathbf{I}\right]$.}
\nomenclature[27]{$\mathbf{G}_i\in\mathbb{R}^{3\times4}$}{$\left[-\mathbf{e}\quad -\Tilde{\mathbf{e}}+e_0\mathbf{I}\right]$.}
\nomenclature[28]{$\mathbf{C}_i\in\mathbb{R}^{3\times3}$}{The rotation matrix that signifies a transformation from the joint definition frame to the $i$\textsuperscript{th} body-fixed frame.}
\nomenclature[29]{$h$}{Time step.}
\printnomenclature

\section*{Acknowledgements}
The authors express their gratitude to Andrey Popov and Reid Gomillion for their assistance in formulating the dynamics and recommending suitable solvers for the differential equations. C. Sandu acknowledges support in part from the Robert E. Hord Jr. Professorship and the Terramechanics, Multibody, and Vehicle Systems Laboratory at Virginia Tech. A. Sandu's work has received support in part from NSF grants ACI-1709727 and CDS\&E-MSS 1953113, DOE grant ASCRDESC0021313, and the Computational Science Laboratory at Virginia Tech. D. Dopico's work was funded by the Spanish Ministry of Science and Innovation (MICINN) under project PID2020-120270GB-C21. The authors extend their thanks to Prof. Ed Haug for providing feedback on this research.

\section*{Statements and Declarations}
The authors have no competing interests related to the work submitted for publication.

%
%

\bibliographystyle{spmpsci}      
\bibliography{bibliography}   

\begin{thebibliography}{10}
\providecommand{\url}[1]{{#1}}
\providecommand{\urlprefix}{URL }
\expandafter\ifx\csname urlstyle\endcsname\relax
  \providecommand{\doi}[1]{DOI~\discretionary{}{}{}#1}\else
  \providecommand{\doi}{DOI~\discretionary{}{}{}\begingroup
  \urlstyle{rm}\Url}\fi

\bibitem{Alappat2020}
Alappat, C., Basermann, A., Bishop, A.R., Fehske, H., Hager, G., Schenk, O.,
  Thies, J., Wellein, G.: A recursive algebraic coloring technique for
  hardware-efficient symmetric sparse matrix-vector multiplication.
\newblock ACM Trans. Parallel Comput. \textbf{7}(3) (2020).
\newblock \doi{10.1145/3399732}.
\newblock \urlprefix\url{https://doi.org/10.1145/3399732}

\bibitem{Alavi2021}
Alavi, A., Dolatabadi, M., Mashhadi, J., Noroozinejad~Farsangi, E.:
  Simultaneous optimization approach for combined control--structural design
  versus the conventional sequential optimization method.
\newblock Structural and Multidisciplinary Optimization \textbf{63}(3),
  1367--1383 (2021).
\newblock \doi{10.1007/s00158-020-02765-3}.
\newblock \urlprefix\url{https://doi.org/10.1007/s00158-020-02765-3}

\bibitem{Allison2014CodesignOA}
Allison, J.T., Han, Z.: Co-design of an active suspension using simultaneous
  dynamic optimization.
\newblock In: Design Automation Conference (2014)

\bibitem{Andersson2019}
Andersson, J.A.E., Gillis, J., Horn, G., Rawlings, J.B., Diehl, M.: Casadi: a
  software framework for nonlinear optimization and optimal control.
\newblock Mathematical Programming Computation \textbf{11}(1), 1--36 (2019).
\newblock \doi{10.1007/s12532-018-0139-4}.
\newblock \urlprefix\url{https://doi.org/10.1007/s12532-018-0139-4}

\bibitem{GMRES2005}
Baker, A.H., Jessup, E.R., Manteuffel, T.: A technique for accelerating the
  convergence of restarted gmres.
\newblock SIAM Journal on Matrix Analysis and Applications \textbf{26}(4),
  962--984 (2005).
\newblock \doi{10.1137/S0895479803422014}.
\newblock \urlprefix\url{https://doi.org/10.1137/S0895479803422014}

\bibitem{baydin2018automatic}
Baydin, A.G., Pearlmutter, B.A., Radul, A.A., Siskind, J.M.: Automatic
  differentiation in machine learning: a survey (2018)

\bibitem{ADIFOR}
Bischof, C., Carle, A., Corliss, G., Griewank, A., Hovland, P.:
  Adifor-generating derivative codes from fortran programs.
\newblock Sci. Program. \textbf{1}(1), 11–29 (1992).
\newblock \doi{10.1155/1992/717832}.
\newblock \urlprefix\url{https://doi.org/10.1155/1992/717832}

\bibitem{Brown2016}
Brown, P., McPhee, J.: {A Continuous Velocity-Based Friction Model for Dynamics
  and Control with Physically Meaningful Parameters}.
\newblock Journal of Computational and Nonlinear Dynamics \textbf{11}(5), 1--6
  (2016).
\newblock \doi{10.1115/1.4033658}

\bibitem{LBFGS_bound}
Byrd, R.H., Lu, P., Nocedal, J., Zhu, C.: A limited memory algorithm for bound
  constrained optimization.
\newblock SIAM Journal on Scientific Computing \textbf{16}(5), 1190--1208
  (1995).
\newblock \doi{10.1137/0916069}.
\newblock \urlprefix\url{https://doi.org/10.1137/0916069}

\bibitem{Alfonso2013_dynamic_response_optimization_using_AD}
Callejo, A.: Dynamic response optimization of vehicles through efficient
  multibody formulations and automatic differentiation techniques.
\newblock Ph.D. thesis, E.T.S.I. Industriales, Universidad Polit\'{e}nica de
  Madrid (2013)

\bibitem{Collins2005}
Collins, S., Ruina, A., Tedrake, R., Wisse, M.: Efficient bipedal robots based
  on passive-dynamic walkers.
\newblock Science \textbf{307}(5712), 1082--1085 (2005).
\newblock \urlprefix\url{http://www.jstor.org/stable/3840156}

\bibitem{Corner2020}
Corner, S., Sandu, A., Sandu, C.: {Adjoint sensitivity analysis of hybrid
  multibody dynamical systems}.
\newblock Multibody System Dynamics \textbf{49}(4), 395--420 (2020).
\newblock \doi{10.1007/s11044-020-09726-0}.
\newblock \urlprefix\url{http://dx.doi.org/10.1007/s11044-020-09726-0}

\bibitem{Corner2019}
Corner, S., Sandu, C., Sandu, A.: {Modeling and sensitivity analysis
  methodology for hybrid dynamical system}.
\newblock Nonlinear Analysis: Hybrid Systems \textbf{31}, 19--40 (2019).
\newblock \doi{10.1016/j.nahs.2018.07.003}.
\newblock \urlprefix\url{https://doi.org/10.1016/j.nahs.2018.07.003}

\bibitem{Dopico2015}
Dopico, D., Zhu, Y., Sandu, A., Sandu, C.: {Direct and adjoint sensitivity
  analysis of ordinary differential equation multibody formulations}.
\newblock Journal of Computational and Nonlinear Dynamics  (2015).
\newblock \doi{10.1115/1.4026492}

\bibitem{dlr144872}
Elmqvist, H., Otter, M., Neumayr, A., Hippmann, G.: Modia - equation based
  modeling and domain specific algorithms.
\newblock In: M.~Sj{\"o}lund, L.~Buffoni, A.~Pop, L.~Ochel (eds.) 14th
  International Modelica Conference, Link{\"o}ping Electronic Conference
  Proceedings 181, pp. 73--86. Link{\"o}ping University Electronic Press
  (2021).
\newblock \urlprefix\url{https://elib.dlr.de/144872/}

\bibitem{FEEHERY199741}
Feehery, W.F., Tolsma, J.E., Barton, P.I.: Efficient sensitivity analysis of
  large-scale differential-algebraic systems.
\newblock Applied Numerical Mathematics \textbf{25}(1), 41--54 (1997).
\newblock \doi{https://doi.org/10.1016/S0168-9274(97)00050-0}.
\newblock
  \urlprefix\url{https://www.sciencedirect.com/science/article/pii/S0168927497000500}

\bibitem{7041347}
Feng, S., Whitman, E., Xinjilefu, X., Atkeson, C.G.: Optimization based full
  body control for the atlas robot.
\newblock In: 2014 IEEE-RAS International Conference on Humanoid Robots, pp.
  120--127 (2014).
\newblock \doi{10.1109/HUMANOIDS.2014.7041347}

\bibitem{Flores2004}
Flores, P., Ambr{\'{o}}sio, J., Claro, J.P.: {Dynamic analysis for planar
  multibody mechanical systems with lubricated joints}.
\newblock Multibody System Dynamics \textbf{12}(1), 47--74 (2004).
\newblock \doi{10.1023/B:MUBO.0000042901.74498.3a}

\bibitem{Flores2006}
Flores, P., Ambrósio, J., Claro, J.C.P., Lankarani, H.M.: {Dynamics of
  Multibody Systems With Spherical Clearance Joints}.
\newblock Journal of Computational and Nonlinear Dynamics \textbf{1}(3),
  240--247 (2006).
\newblock \doi{10.1115/1.2198877}.
\newblock \urlprefix\url{https://doi.org/10.1115/1.2198877}

\bibitem{FLORES2023105305}
Flores, P., Ambrósio, J., Lankarani, H.M.: Contact-impact events with friction
  in multibody dynamics: Back to basics.
\newblock Mechanism and Machine Theory \textbf{184}, 105305 (2023).
\newblock \doi{https://doi.org/10.1016/j.mechmachtheory.2023.105305}.
\newblock
  \urlprefix\url{https://www.sciencedirect.com/science/article/pii/S0094114X23000782}

\bibitem{Gonthier2004}
Gonthier, Y., McPhee, J., Lange, C., Piedb{\oe}uf, J.C.: A regularized contact
  model with asymmetric damping and dwell-time dependent friction.
\newblock Multibody System Dynamics \textbf{11}(3), 209--233 (2004).
\newblock \doi{10.1023/B:MUBO.0000029392.21648.bc}.
\newblock \urlprefix\url{https://doi.org/10.1023/B:MUBO.0000029392.21648.bc}

\bibitem{gowda2021high}
Gowda, S., Ma, Y., Cheli, A., Gwozdz, M., Shah, V.B., Edelman, A., Rackauckas,
  C.: High-performance symbolic-numerics via multiple dispatch.
\newblock arXiv preprint arXiv:2105.03949  (2021)

\bibitem{gowda2019sparsity}
Gowda, S., Ma, Y., Churavy, V., Edelman, A., Rackauckas, C.: Sparsity
  programming: Automated sparsity-aware optimizations in differentiable
  programming.
\newblock In: Program Transformations for ML Workshop at NeurIPS 2019 (2019).
\newblock \urlprefix\url{https://openreview.net/forum?id=rJlPdcY38B}

\bibitem{Griewank2008}
Griewank, A., Walther, A.: Evaluating Derivatives, second edn.
\newblock Society for Industrial and Applied Mathematics (2008).
\newblock \doi{10.1137/1.9780898717761}.
\newblock
  \urlprefix\url{https://epubs.siam.org/doi/abs/10.1137/1.9780898717761}

\bibitem{hairer1999}
Hairer, E., Wanner, G.: Stiff differential equations solved by radau methods.
\newblock Journal of Computational and Applied Mathematics \textbf{111}(1),
  93--111 (1999).
\newblock \doi{https://doi.org/10.1016/S0377-0427(99)00134-X}.
\newblock
  \urlprefix\url{https://www.sciencedirect.com/science/article/pii/S037704279900134X}

\bibitem{hairer2010solving}
Hairer, E., Wanner, G.: Solving Ordinary Differential Equations II: Stiff and
  Differential-Algebraic Problems.
\newblock Springer Series in Computational Mathematics. Springer Berlin
  Heidelberg (2010)

\bibitem{Haug2021}
Haug, E.: Computer Aided Kinematics and Dynamics of Mechanical Systems Vol II:
  Modern Methods.
\newblock Research Gate (2021)

\bibitem{Haug1989}
Haug, E.J.: {Computer aided kinematics and dynamics of mechanical systems,
  volume 1: Basic methods}.
\newblock Pearson College Div, Massachusetts (1989).
\newblock \doi{10.1016/0278-6125(92)90050-p}

\bibitem{Haug2018b}
Haug, E.J.: {Simulation of spatial multibody systems with friction}.
\newblock Mechanics Based Design of Structures and Machines \textbf{46}(3),
  347--375 (2018).
\newblock \doi{10.1080/15397734.2017.1377086}.
\newblock \urlprefix\url{https://doi.org/10.1080/15397734.2017.1377086}

\bibitem{hindmarsh1983odepack}
Hindmarsh, A.C.: Odepack, a systemized collection of ode solvers.
\newblock Scientific computing  (1983)

\bibitem{Hindmarsh2005}
Hindmarsh, A.C., Brown, P.N., Grant, K.E., Lee, S.L., Serban, R., Shumaker,
  D.E., Woodward, C.S.: Sundials: Suite of nonlinear and differential/algebraic
  equation solvers.
\newblock ACM Trans. Math. Softw. \textbf{31}(3), 363–396 (2005).
\newblock \doi{10.1145/1089014.1089020}.
\newblock \urlprefix\url{https://doi.org/10.1145/1089014.1089020}

\bibitem{Holland1992}
Holland, J.H.: Adaptation in Natural and Artificial Systems: An Introductory
  Analysis with Applications to Biology, Control and Artificial Intelligence.
\newblock MIT Press, Cambridge, MA, USA (1992)

\bibitem{Garcia2013}
Garc{\'i}a~de Jal{\'o}n, J., Guti{\'e}rrez-L{\'o}pez, M.D.: Multibody dynamics
  with redundant constraints and singular mass matrix: existence, uniqueness,
  and determination of solutions for accelerations and constraint forces.
\newblock Multibody System Dynamics \textbf{30}(3), 311--341 (2013).
\newblock \doi{10.1007/s11044-013-9358-7}.
\newblock \urlprefix\url{https://doi.org/10.1007/s11044-013-9358-7}

\bibitem{Fraczek2011}
Janusz, F., Wojtyra, M.: {On the unique solvability of a direct dynamics
  problem for mechanisms with redundant constraints and Coulomb friction in
  joints}.
\newblock Mechanism and Machine Theory \textbf{46}(3), 312--334 (2011).
\newblock \doi{10.1016/j.mechmachtheory.2010.11.003}

\bibitem{NewtonKrylov}
Kelley, C.T.: Solving Nonlinear Equations with Newton's Method, chap.~3, p.
  57–83.
\newblock Society for Industrial and Applied Mathematics (2003).
\newblock \doi{10.1137/1.9780898718898}.
\newblock
  \urlprefix\url{https://epubs.siam.org/doi/abs/10.1137/1.9780898718898}

\bibitem{Kennedy1995}
Kennedy, J., Eberhart, R.: Particle swarm optimization.
\newblock In: Proceedings of ICNN'95 - International Conference on Neural
  Networks, vol.~4, pp. 1942--1948 vol.4 (1995).
\newblock \doi{10.1109/ICNN.1995.488968}

\bibitem{Kim_2021}
Kim, S., Ji, W., Deng, S., Ma, Y., Rackauckas, C.: Stiff neural ordinary
  differential equations.
\newblock Chaos: An Interdisciplinary Journal of Nonlinear Science
  \textbf{31}(9) (2021).
\newblock \doi{10.1063/5.0060697}

\bibitem{KNOLL2004357}
Knoll, D., Keyes, D.: Jacobian-free newton–krylov methods: a survey of
  approaches and applications.
\newblock Journal of Computational Physics \textbf{193}(2), 357--397 (2004).
\newblock \doi{https://doi.org/10.1016/j.jcp.2003.08.010}.
\newblock
  \urlprefix\url{https://www.sciencedirect.com/science/article/pii/S0021999103004340}

\bibitem{rigidbodydynamicsjl}
Koolen, T., contributors: Rigidbodydynamics.jl (2016).
\newblock \urlprefix\url{https://github.com/JuliaRobotics/RigidBodyDynamics.jl}

\bibitem{LI2000131}
Li, S., Petzold, L.: Software and algorithms for sensitivity analysis of
  large-scale differential algebraic systems.
\newblock Journal of Computational and Applied Mathematics \textbf{125}(1),
  131--145 (2000).
\newblock \doi{https://doi.org/10.1016/S0377-0427(00)00464-7}.
\newblock
  \urlprefix\url{https://www.sciencedirect.com/science/article/pii/S0377042700004647}.
\newblock Numerical Analysis 2000. Vol. VI: Ordinary Differential Equations and
  Integral Equations

\bibitem{Ma2021_comparison_of_AD_in_DE}
Ma, Y., Dixit, V., Innes, M.J., Guo, X., Rackauckas, C.: A comparison of
  automatic differentiation and continuous sensitivity analysis for derivatives
  of differential equation solutions.
\newblock In: 2021 IEEE High Performance Extreme Computing Conference (HPEC),
  pp. 1--9 (2021).
\newblock \doi{10.1109/HPEC49654.2021.9622796}

\bibitem{ma2021modelingtoolkit}
Ma, Y., Gowda, S., Anantharaman, R., Laughman, C., Shah, V., Rackauckas, C.:
  Modelingtoolkit: A composable graph transformation system for equation-based
  modeling (2021)

\bibitem{Maciag2020_hamiltonian_sensitivity}
Maci\k{a}g, P., Malczyk, P., Fr\k{a}czek, J.: Hamiltonian direct
  differentiation and adjoint approaches for multibody system sensitivity
  analysis.
\newblock International Journal for Numerical Methods in Engineering
  \textbf{121}(22), 5082--5100 (2020).
\newblock \doi{https://doi.org/10.1002/nme.6512}.
\newblock
  \urlprefix\url{https://onlinelibrary.wiley.com/doi/abs/10.1002/nme.6512}

\bibitem{Marques2019}
Marques, F., Flores, P., Claro, J.C.P., Lankarani, H.M.: {Modeling and analysis
  of friction including rolling effects in multibody dynamics: a review}.
\newblock Multibody System Dynamics \textbf{45}(2), 223--244 (2019).
\newblock \doi{10.1007/s11044-018-09640-6}.
\newblock \urlprefix\url{http://dx.doi.org/10.1007/s11044-018-09640-6}

\bibitem{McCourt2015}
McCourt, M., Smith, B., Zhang, H.: Sparse matrix-matrix products executed
  through coloring.
\newblock SIAM Journal on Matrix Analysis and Applications \textbf{36}(1),
  90--109 (2015).
\newblock \doi{10.1137/13093426X}.
\newblock \urlprefix\url{https://doi.org/10.1137/13093426X}

\bibitem{RePEc:mtp:titles:0262633094}
Miranda, M.J., Fackler, P.L.: {Applied Computational Economics and Finance},
  \emph{MIT Press Books}, vol.~1.
\newblock The MIT Press (2004).
\newblock \urlprefix\url{https://ideas.repec.org/b/mtp/titles/0262633094.html}

\bibitem{osti_6997568}
More, J.J., Garbow, B.S., Hillstrom, K.E.: User guide for {MINPACK}-1. [{I}n
  {FORTRAN}].
\newblock Web  (1980).
\newblock \doi{10.2172/6997568}

\bibitem{Muchnick1997}
Muchnick, S.S..: Advanced compiler design and implementation.
\newblock Morgan Kaufmann Publishers, San Francisco, Calif. (1997).
\newblock \urlprefix\url{http://catdir.loc.gov/catdir/toc/els032/97013063.html}

\bibitem{simultaneousDC1}
Nakka, S.K.S., Alexander-Ramos, M.J.: Simultaneous combined optimal design and
  control formulation for aircraft hybrid-electric propulsion systems.
\newblock Journal of Aircraft \textbf{58}(1), 53--62 (2021).
\newblock \doi{10.2514/1.C035678}.
\newblock \urlprefix\url{https://doi.org/10.2514/1.C035678}

\bibitem{Nelder1965}
Nelder, J.A., Mead, R.: {A Simplex Method for Function Minimization}.
\newblock The Computer Journal \textbf{7}(4), 308--313 (1965).
\newblock \doi{10.1093/comjnl/7.4.308}.
\newblock \urlprefix\url{https://doi.org/10.1093/comjnl/7.4.308}

\bibitem{NoceWrig06}
Nocedal, J., Wright, S.J.: Numerical Optimization, 2e edn.
\newblock Springer, New York, NY, USA (2006)

\bibitem{Orden2005}
Orden, J.C.G.: Analysis of joint clearances in multibody systems.
\newblock Multibody System Dynamics \textbf{13}(4), 401--420 (2005).
\newblock \doi{10.1007/s11044-005-3989-2}.
\newblock \urlprefix\url{https://doi.org/10.1007/s11044-005-3989-2}

\bibitem{Pennestri2007}
Pennestr{\`{i}}, E., Valentini, P.P., Vita, L.: {Multibody dynamics simulation
  of planar linkages with Dahl friction}.
\newblock Multibody System Dynamics \textbf{17}(4), 321--347 (2007).
\newblock \doi{10.1007/s11044-007-9047-5}

\bibitem{doi:10.1137/0904010}
Petzold, L.: Automatic selection of methods for solving stiff and nonstiff
  systems of ordinary differential equations.
\newblock SIAM Journal on Scientific and Statistical Computing \textbf{4}(1),
  136--148 (1983).
\newblock \doi{10.1137/0904010}.
\newblock \urlprefix\url{https://doi.org/10.1137/0904010}

\bibitem{Petzold2006}
Petzold, L., Li, S., Cao, Y., Serban, R.: Sensitivity analysis of
  differential-algebraic equations and partial differential equations.
\newblock Computers \& Chemical Engineering \textbf{30}(10), 1553--1559 (2006).
\newblock \doi{https://doi.org/10.1016/j.compchemeng.2006.05.015}.
\newblock
  \urlprefix\url{https://www.sciencedirect.com/science/article/pii/S0098135406001487}.
\newblock Papers form Chemical Process Control VII

\bibitem{osti_5882821}
Petzold, L.R.: Description of {DASSL}: a differential/algebraic system solver.
\newblock Web  (1982).
\newblock \urlprefix\url{https://www.osti.gov/biblio/5882821}

\bibitem{PIKULINSKI2021104473}
Pikuli\'{n}ski, M., Malczyk, P.: Adjoint method for optimal control of
  multibody systems in the hamiltonian setting.
\newblock Mechanism and Machine Theory \textbf{166}, 104473 (2021).
\newblock \doi{https://doi.org/10.1016/j.mechmachtheory.2021.104473}.
\newblock
  \urlprefix\url{https://www.sciencedirect.com/science/article/pii/S0094114X21002317}

\bibitem{ADofDEsolvers}
Rackauckas, C.: Direct automatic differentiation of (differential equation)
  solvers vs analytical adjoints: Which is better?
\newblock
  \url{https://www.stochasticlifestyle.com/direct-automatic-differentiation-of-solvers-vs-analytical-adjoints-which-is-better/}
  (2022)

\bibitem{rackauckas2021composing}
Rackauckas, C., Anantharaman, R., Edelman, A., Gowda, S., Gwozdz, M., Jain, A.,
  Laughman, C., Ma, Y., Martinuzzi, F., Pal, A., Rajput, U., Saba, E., Shah,
  V.B.: Composing modeling and simulation with machine learning in julia (2021)

\bibitem{rackauckas2021universal}
Rackauckas, C., Ma, Y., Martensen, J., Warner, C., Zubov, K., Supekar, R.,
  Skinner, D., Ramadhan, A., Edelman, A.: Universal differential equations for
  scientific machine learning (2021)

\bibitem{radhakrishnan1993description}
Radhakrishnan, K., Hindmarsh, A.C.: Description and use of lsode, the livermore
  solver for ordinary differential equations.
\newblock Tech. rep., Lawrence Livermore National Laboratory (1993)

\bibitem{RevelsLubinPapamarkou2016_ForwardADJulia}
{Revels}, J., {Lubin}, M., {Papamarkou}, T.: Forward-mode automatic
  differentiation in {J}ulia.
\newblock arXiv:1607.07892 [cs.MS]  (2016).
\newblock \urlprefix\url{https://arxiv.org/abs/1607.07892}

\bibitem{Rothwell2017}
Rothwell, A.: Numerical Methods for Unconstrained Optimization, pp. 83--106.
\newblock Springer International Publishing, Cham (2017).
\newblock \doi{10.1007/978-3-319-55197-5\_4}.
\newblock \urlprefix\url{https://doi.org/10.1007/978-3-319-55197-5\_4}

\bibitem{Serban1997_AugI1DAE}
Serban, R., Negrut, D., Haug, E.J., Potra, F.A.: A topology-based approach for
  exploiting sparsity in multibody dynamics in cartesian formulation.
\newblock Mechanics of Structures and Machines \textbf{25}(3), 379--396 (1997).
\newblock \doi{10.1080/08905459708905295}.
\newblock \urlprefix\url{https://doi.org/10.1080/08905459708905295}

\bibitem{steinebach1995order}
Steinebach, G.: Order reduction of ROW methods for DAEs and method of lines
  applications.
\newblock Preprint. Techn. Hochsch., Fachbereich Mathematik (1995).
\newblock \urlprefix\url{https://books.google.com/books?id=6TRDHQAACAAJ}

\bibitem{Steinebach2023}
Steinebach, G.: Construction of rosenbrock--wanner method rodas5p and numerical
  benchmarks within the julia differential equations package.
\newblock BIT Numerical Mathematics \textbf{63}(2), 27 (2023).
\newblock \doi{10.1007/s10543-023-00967-x}.
\newblock \urlprefix\url{https://doi.org/10.1007/s10543-023-00967-x}

\bibitem{Tian2018}
Tian, Q., Flores, P., Lankarani, H.M.: {A comprehensive survey of the
  analytical, numerical and experimental methodologies for dynamics of
  multibody mechanical systems with clearance or imperfect joints}.
\newblock Mechanism and Machine Theory \textbf{122}, 1--57 (2018).
\newblock \doi{10.1016/j.mechmachtheory.2017.12.002}.
\newblock \urlprefix\url{https://doi.org/10.1016/j.mechmachtheory.2017.12.002}

\bibitem{Alvaro_ALI3P_direct_sensitivity}
Varela, A.L., Dopico, D.D., Fern\'{a}ndez, A.L.: Augmented lagrangian index-3
  semi-recursive formulations with projections. direct sensitivity analysis
  (2023).
\newblock \doi{10.21203/rs.3.rs-2687084/v1}.
\newblock \urlprefix\url{https://doi.org/10.21203/rs.3.rs-2687084/v1}

\bibitem{Verulkar2022}
Verulkar, A., Sandu, C., Dopico, D., Sandu, A.: Computation of direct
  sensitivities of spatial multibody systems with joint friction.
\newblock Journal of Computational and Nonlinear Dynamics \textbf{17}(7),
  071006 (2022).
\newblock \doi{10.1115/1.4054110}.
\newblock \urlprefix\url{https://doi.org/10.1115/1.4054110}

\bibitem{verulkaroptimal}
Verulkar, A., Sandu, C., Sandu, A., Dopico, D.: Optimal control of multibody
  systems in reduced space {ODE} formulation.
\newblock In: {IUTAM} Symposium on Optimal Design and Control of Multibody
  Systems (2022)

\bibitem{IPOPT2006}
W{\"a}chter, A., Biegler, L.: On the implementation of an interior-point filter
  line-search algorithm for large-scale nonlinear programming.
\newblock Mathematical Programming \textbf{106}(1), 25--57 (2006).
\newblock \doi{10.1007/s10107-004-0559-y}.
\newblock Copyright: Copyright 2008 Elsevier B.V., All rights reserved.

\bibitem{wanner1996solving}
Wanner, G., Hairer, E.: Solving ordinary differential equations II, vol. 375.
\newblock Springer Berlin Heidelberg New York (1996)

\bibitem{LuGre}
Canudas~de Wit, C., Olsson, H., Astrom, K., Lischinsky, P.: A new model for
  control of systems with friction.
\newblock IEEE Transactions on Automatic Control \textbf{40}(3), 419--425
  (1995).
\newblock \doi{10.1109/9.376053}

\bibitem{MBSVT_IDETC2014}
Zhu, Y., Dopico, D., Sandu, C., Sandu, A.: {MBSVT: Software for Modeling,
  Sensitivity Analysis, and Optimization of Multibody Systems at Virginia
  Tech}.
\newblock In: International Design Engineering Technical Conferences and
  Computers and Information in Engineering Conference, vol. Volume 7: 2nd
  Biennial International Conference on Dynamics for Design; 26th International
  Conference on Design Theory and Methodology, p. V007T05A001 (2014).
\newblock \doi{10.1115/DETC2014-34084}.
\newblock \urlprefix\url{https://doi.org/10.1115/DETC2014-34084}

\bibitem{AKESSON20101737}
Åkesson, J., Årzén, K.E., Gäfvert, M., Bergdahl, T., Tummescheit, H.:
  Modeling and optimization with optimica and jmodelica.org—languages and
  tools for solving large-scale dynamic optimization problems.
\newblock Computers \& Chemical Engineering \textbf{34}(11), 1737--1749 (2010).
\newblock \doi{https://doi.org/10.1016/j.compchemeng.2009.11.011}.
\newblock
  \urlprefix\url{https://www.sciencedirect.com/science/article/pii/S009813540900283X}

\end{thebibliography}
\end{document}